\global\def\draftcontrol{0}

   \def\versionno{ kt transport -- draft   }

\catcode`\@=11

\expandafter\ifx\csname draftcontrol\endcsname\relax\global\def\draftcontrol{0}
\fi

{\count255=\time\divide\count255 by 60
\xdef\hourmin{\number\count255}
\multiply\count255 by-60\advance\count255 by\time
\xdef\hourmin{\hourmin:\ifnum\count255<10 0\fi\the\count255}}
\def\draftdate{\number\month/\number\day/\number\year\ \ \ \hourmin }

\newcommand\makepapertitle{\par
  \begingroup
    \renewcommand\thefootnote{\@fnsymbol\c@footnote}%
    \def\@makefnmark{\rlap{\@textsuperscript{\normalfont\@thefnmark}}}%
    \long\def\@makefntext##1{\parindent 1em\noindent
            \hb@xt@1.8em{%
                \hss\@textsuperscript{\normalfont\@thefnmark}}##1}%
     \newpage
     \global\@topnum\z@   
     \@makepapertitle
     \thispagestyle{empty}\@thanks
  \endgroup
  \setcounter{footnote}{0}%
  \global\let\thanks\relax
  \global\let\makepapertitle\relax
  \global\let\@makepapertitle\relax
  \global\let\@thanks\@empty
  \global\let\@author\@empty
  \global\let\@date\@empty
  \global\let\@title\@empty
  \global\let\title\relax
  \global\let\author\relax
  \global\let\date\relax
  \global\let\and\relax
  \def\version{\let\version\@version\@gobble}
}
\def\@makepapertitle{%
  \newpage
   \ifnum\draftcontrol=1 {}
   \version\versionno
   \vskip 3em%
   \else
   \hfill\hbox to 3cm {\parbox{4cm}{\@pubnum}\hss}%
   \vskip 3em%
   \fi
   \begin{center}%
   \let \footnote \thanks
     {\LARGE {\@title}}%
     \vskip 1.5em%
     {\normalsize
       \lineskip .5em%
       \begin{tabular}[t]{c}%
         \@author
       \end{tabular}\par}%
     \vskip 1.5em%
     {\@bstract}%
     \end{center}%
     \vskip 1.5em 
     \@date%
   \par
}

\gdef\@pubnum{}
\def\pubnum#1{%
  \gdef\@pubnum{#1}}

\gdef\@bstract{}
\def\Abstract#1{%
  \gdef\@bstract{%
   \parbox{\textwidth-0pc}{%
   \centerline{\bf Abstract}\penalty1000
   \noindent
   \renewcommand\baselinestretch{1.0}
   {#1}}}
}

\def\ps@paper{\let\@mkboth\@gobbletwo%
     \ifnum\draftcontrol=1
        \def\@oddfoot{\hbox to \textwidth{\tiny \versionno \hfil\tiny\draftdate}%
        \hskip -\textwidth \hbox to \textwidth{\hfil\rm\thepage\hfil}}%
     \else\def\@oddfoot{\hbox to \textwidth{\hfil\rm\thepage\hfil}}
     \fi
     \let\@evenfoot\@oddfoot
}

\def\body{\clearpage
          \pagestyle{paper}
        }

\def\@version#1{\ifnum\draftcontrol=1
\typeout{}\typeout{#1}\typeout{}
\vskip3mm\centerline{\hbox{\fbox{\normalsize{\tt DRAFT -- #1 -- }
                   {\draftdate}}}}\vskip3mm
\fi}
\let\version\@version
\long\def\eqlabel#1{\ifnum\draftcontrol=1
                    \tag@false  
                    \tag*{(\theequation) \hbox to -0.2cm{\hspace{0cm}\small{#1}\hss}}
                    \refstepcounter{equation} 
                    \edef\@currentlabel{\theequation}
                    \ltx@label{#1}          
                    \else
                    \label{#1}
                    \fi
                    }
\let\st@bibitem\@bibitem
\let\st@lbibitem\@lbibitem
\ifnum\draftcontrol=1
  \def\@bibitem#1{%
    \st@bibitem{#1}\a@@label{#1}\ignorespaces}
  \def\@lbibitem[#1]#2{%
    \st@lbibitem[#1]{#2}\a@@label{#2}\ignorespaces}
  \def\a@@label#1{%
    \gdef\a@lab{\smash{\normalfont\small#1}}
    \ifvmode
      \if@inlabel
        \global\setbox\@labels\hbox{%
          \llap{\a@lab\let\a@lab\relax
                \kern\@totalleftmargin\kern\marginparsep}%
          \box\@labels}%
      \fi
    \fi}
\fi

\documentclass[12pt,letterpaper]{article}

\usepackage{amsmath,amssymb,array,calc,rotating,epsfig,psfrag}
\usepackage[nosort]{cite}

\ifnum\draftcontrol=1
\fi

\renewcommand\baselinestretch{1.25}
\renewcommand\section{\@startsection {section}{1}{\z@}%
                                   {-3.5ex \@plus -1ex \@minus -.2ex}%
                                   {2.3ex \@plus.2ex}%
                                   {\normalfont\large\bfseries}}
\renewcommand\subsection{\@startsection{subsection}{2}{\z@}%
                                   {-3.25ex\@plus -1ex \@minus -.2ex}%
                                   {1.5ex \@plus .2ex}%
                                   {\normalfont\normalsize\bfseries}}
\renewcommand\subsubsection{\@startsection{subsubsection}{3}{\z@}%
                                   {-3.25ex\@plus -1ex \@minus -.2ex}%
                                   {1.5ex \@plus .2ex}%
                                   {\normalfont\normalsize\it}}
\renewcommand\paragraph{\@startsection{paragraph}{4}{\z@}%
                                   {-3.25ex\@plus -1ex \@minus -.2ex}%
                                   {1.5ex \@plus .2ex}%
                                   {\normalfont\normalsize\bf}}

\def\ie{{\it i.e.}}

\def\revise#1       {\raisebox{-0em}{\rule{3pt}{1em}}%
                     \marginpar{\raisebox{.5em}{\vrule width3pt\
                     \vrule width0pt height 0pt depth0.5em
                     \hbox to 0cm{\hspace{0cm}{%
                     \parbox[t]{4em}{\raggedright\footnotesize{#1}}}\hss}}}}

\newcommand\nxt[1]  {\\\fnxt#1}

\def\calc         {{\cal C}}

\def\calf         {{\cal F}}

\def\calk         {{\cal K}}

\def\calm         {{\cal M}}
\def\caln         {{\cal N}}
\def\calo         {{\cal O}}
\def\calp         {{\cal P}}

\def\del          {\partial}

\def\ee           {{\rm e}}

\def\sqr#1#2{{\vcenter{\vbox{\hrule height.#2pt  
 \hbox{\vrule width.#2pt height#1pt \kern#1pt
 \vrule width.#2pt}\hrule height.#2pt}}}}
\def\square{%
  \mathop{\mathchoice{\sqr{12}{15}}{\sqr{9}{12}}{\sqr{6.3}{9}}{\sqr{4.5}{9}}}}

\newcommand{\ft}[2]{{\textstyle{\frac{#1}{#2}}}}

\newcommand{\qq}{\mathfrak{q}}
\newcommand{\ww}{\mathfrak{w}}

\def\om{\Omega}

\def\dd{\delta}

\def\ee{\epsilon}
\def\hR{\hat{R}}

\def\hg{\hat{g}}
\def\hx{\hat{x}}
\def\hy{\hat{y}}
\def\hz{\hat{z}}

\def\w{\omega}
\def\hh{\hat{h}}

\def\aa1{\phi}
\def\cc1{\psi}
\def\k{\calk}
\def\dilog{\rm dilog}
\def\b{\beta}
\def\G{\Gamma}
\def\C{\calc}
\def\hc{\hat{\calc}}
\def\hpsi{\hat{\psi}}

\catcode`\@=12

\begin{document}


\title{Transport properties of cascading gauge theories}

\pubnum{%
UWO-TH-05/08\\
hep-th/0509083}
\date{September 2005}

\author{Alex Buchel\\[0.4cm]
\it Perimeter Institute for Theoretical Physics\\
  \it Waterloo, Ontario N2J 2W9, Canada\\[0.2cm]
  \it Department of Applied Mathematics, University of Western Ontario\\
  \it London, Ontario N6A 5B7, Canada
 }

\Abstract{
Cascading gauge theories of Klebanov {\it et.al.} provide a model
within a framework of gauge theory/string theory duality for a four
dimensional non-conformal gauge theory with a spontaneously generated
mass scale. Using the dual supergravity description we study sound
wave propagation in strongly coupled cascading gauge theory plasma.
We analytically compute the speed of sound and the bulk viscosity of
cascading gauge theory plasma at a temperature much larger than the
strong coupling scale of the theory. The sound wave dispersion
relation is obtained from the hydrodynamic pole in the stress-energy
tensor two-point correlation function.  The speed of sound extracted
from the pole of the correlation function agrees with its value
computed in [hep-th/0506002] using the equation of state. We find that
the bulk viscosity of the hot cascading gauge theory plasma is
non-zero at the leading order in the deviation from conformality.
}


\makepapertitle

\body

\version\versionno

\section{Introduction}
The correspondence between gauge theories and  string theory of Maldacena \cite{m9711,m2}
has become a valuable tool in analyzing near-equilibrium dynamics of strongly coupled 
gauge theory plasma \cite{ss,hs,ne1,ne2,ne3,ne4,ne5,kss,bh1,bl1,kss1,bls,bh2,set,bbs}. The research in 
this direction is primarily motivated
by its potential application for the hydrodynamic description of the QCD quark-gluon plasma 
believed to be produced in heavy ion collision experiments at RHIC \cite{r1,r2,r3}. 
 
Previously, the dual supergravity computations \cite{ne4} were shown to reproduce the expected 
dispersion relation for sound waves in 
strongly coupled $\caln=4$ supersymmetric Yang-Mills (SYM) theory plasma
\begin{equation}
\w(q)=v_s q -i\ \frac{2q^2}{3 T }\ \frac{\eta}{s}\ \left(1+\frac{3\zeta}{4\eta}\right)\,, 
\eqlabel{dispertion}
\end{equation} 
where $v_s, \eta, s, \zeta$ are the plasma sound speed, shear viscosity, entropy density, and bulk viscosity correspondingly. 
Conformal symmetry of the $\caln=4$ gauge theory insures that 
\begin{equation}
v_s=\frac{1}{\sqrt{3}}\,,\qquad \zeta=0 \,. 
\eqlabel{n4dis}
\end{equation}
Non-conformal gauge theories, and QCD in particular, is expected to have nonvanishing bulk viscosity. 
Additionally, the presence of a scale in a gauge theory breaks conformal invariance 
responsible for  speed of sound expression in Eq.~\eqref{n4dis}. To see this note that in conformal theories 
the energy density and the pressure are related as follows $\ee=3P$, thus, 
$v_s^2=\frac{\del P}{\del \ee}=\frac{1}{3}$.

The first computation for  the speed of sound and its attenuation in 
strongly coupled non-conformal four dimensional gauge theories from the dual supergravity
perspective were reported in \cite{bbs}. Specifically, the model considered there was a mass 
deformation of the $\caln=4$ Yang-Mills theory by giving  identical masses $m_b$ to bosonic
components of the two SYM chiral multiples and identical masses $m_f$ to fermionic components of the 
same chiral multiples. Generically $m_f\ne m_b$ and the supersymmetry is completely broken.
For a special case $m_f=m_b$ such a deformation leaves eight supersymmetries unbroken, 
and the model is usually referred to as $\caln=2^*$ gauge theory \cite{w,pw,bpp,j}. 
Using finite temperature  gauge/gravity correspondence  for this mass deformed $\caln=4$ 
SYM \cite{bl} the speed of sound and the ratio of the shear to bulk
viscosity was found, respectively,
\begin{equation}
v_s= \frac{1}{\sqrt{3}} \left( 1-\frac{ \left[\Gamma\left(\ft
34\right)\right]^4}{3\pi^4}\
 \left(\frac{m_f}{T}\right)^2
-\frac{1}{18\pi^4}\ \left(\frac{m_b}{T}\right)^4 +\cdots\right)\,,
\eqlabel{csa}
\end{equation}
\begin{equation}
\frac{\zeta}{\eta} = \beta_f^\Gamma \, \frac{
\left[\Gamma\left(\ft 34\right)\right]^4}{3\pi^3}\
 \left(\frac{m_f}{T}\right)^2 +  \frac{\beta_b^\Gamma}{432\pi^2}\
\left(\frac{m_b}{T}\right)^4 +\cdots\,, \eqlabel{csax}
\end{equation}
where $\beta_f^\Gamma \approx 0.9672$,
 $\beta_b^\Gamma \approx 8.001$, and
 the ellipses
 denote higher order terms in $m_f/T$ and $m_b/T$.
From the dependence in Eqs.~\eqref{csa}, \eqref{csax} it follows that at
least
 in the high temperature regime the
ratio of bulk viscosity to shear viscosity is proportional to the
deviation of the speed of sound squared from its value in
conformal theory,
\begin{equation}
\frac{\zeta}{\eta} \simeq - \kappa \, \left(
v_s^2-\frac{1}{3}\right)\,,
\eqlabel{proporo}
\end{equation}
where  $\kappa=3\pi\beta_f^\Gamma/2 \approx  4.558$
for $m_b=0$, and $\kappa=\pi^2\beta_b^\Gamma/16 
\approx 4.935$ for $m_f= 0$.

In this paper  we extend analysis of the transport properties in strongly coupled non-conformal gauge theory 
plasma to theories with spontaneously generated mass scale. More precisely, we analytically compute 
the speed of sound and its attenuation in 'cascading gauge theories'  \cite{kn,kt,ks} (see \cite{ktreview} for a 
recent review)  at temperature 
much higher than the deconfinement and the chiral symmetry breaking scales of the theory. 
The relevant finite temperature deformations of the theory were discussed in \cite{bhkt1,bhkt2,bhkt3}; 
the holographic renormalization of the cascading gauge theories was explained in \cite{aby}.
Moreover, in \cite{aby}  the equation of state describing hot cascading gauge theory plasma 
was obtained from which  the speed of sound was predicted to be 
\begin{equation}
v_s^2=\frac{\del \calp}{\del\cal\epsilon}=\frac{\frac{\del\calp}{\del
T}}{\frac{\del\cal\epsilon}{\del T}} =\frac 13
-\frac{2}{9\ln\frac{T}{\Lambda}}+\cdots\,,
\eqlabel{ktsound}
\end{equation}
where $\Lambda$ is the strong coupling scale of the theory and  ellipses denote subdominant terms for $T\gg \Lambda$.
Here, we extract the dispersion relation \eqref{dispertion} for the cascading gauge theory plasma
from the pole of the thermal two-point function of certain components of the stress-energy tensor in the 
hydrodynamic approximation, \ie, in the regime where energy and momentum are small in comparison with the 
inverse thermal wavelength $(\omega/T\ll 1, q/T\ll 1)$, and at high temperature $T\gg \Lambda$.
The latter computation is equivalent \cite{set} to determining the dispersion for the lowest quasinormal mode in 
the corresponding black brane  geometry (in our case \cite{bhkt3}). 
The general prescription for computing the quasinormal modes introduced  in 
\cite{set} has been  applied recently to a variety of gauge/gravity  duality examples \cite{qu1,bbs,qu2}.

As the computations are rather  technical, we begin by summarizing our results in the next section.  
In section 3 we review five-dimensional effective supergravity description dual to chirally symmetric 
phase of the cascading gauge theory \cite{aby} and   the explicit analytic construction of the background
black brane geometry \cite{bhkt3} dual to thermal cascading gauge theory at temperature much larger than its strong coupling 
scale. In section 4 we study fluctuations of the corresponding black brane geometry dual to a sound 
wave mode of the cascading gauge theory plasma. We  introduce gauge invariant 
fluctuations and obtain their equations of motion. These equations of motion are valid beyond the 
hydrodynamic approximation, and for arbitrary temperature  (as long as it is higher than the chiral symmetry 
breaking scale of the cascading gauge theory). In section 5 we derive and solve fluctuation equations 
in the hydrodynamic limit and at temperatures much higher than the cascading gauge theory strong coupling scale. 
Imposing Dirichlet condition on the gauge invariant fluctuations at the boundary of the 
background black brane geometry determines \cite{set} 
the dispersion relation for the lowest quasinormal frequency \eqref{dispertion}.
Using the universality of the shear viscosity to entropy density ratio in strongly coupled 
gauge theory \cite{bl1,kss1,bh2},   computed dispersion relation  can further be used to 
evaluate the ratio of bulk to shear viscosities.   
Some computational details are delegated to Appendices A, B and C.

\section{Summary of results}
Cascading gauge theory at a given high-energy scale resembles $\caln=1$ supersymmetric $SU(K_*)\times SU(K_*+P)$ 
gauge theory with two bifundamental and two anti-fundamental chiral superfields and certain superpotential,
which is quartic in superfields. 
Unlike ordinary quiver gauge theories, an 'effective rank' of  cascading gauge theories depends on an energy scale 
at which the theory is probed \cite{bhkt1,bhkt3,aby}
\begin{equation}
K_*\equiv K_*(E)\sim 2 P^2\ \ln\frac{E}{\Lambda}\,,\qquad E\gg \Lambda\,.
\eqlabel{kdep}
\end{equation}
At a given temperature $T$ cascading gauge theory is probed at energy scale $E\sim T$, 
and as $T\gg \Lambda$, $K_*(T)\gg P^2$. In this regime the thermal properties of the theory \cite{bhkt3,aby} are very similar to 
those of the $\caln=1$ $SU(K_*)\times SU(K_*)$ superconformal gauge theory of Klebanov and Witten \cite{kw},
with 
\begin{equation}
\dd_{cascade}\equiv \frac{P^2}{K_*} 
\eqlabel{defcas}  
\end{equation}
being the deformation parameter. Clearly, at the temperature increases, $\dd_{cascade}$ becomes smaller and smaller. 
From the equation of state for the cascading gauge theory  one obtains the speed of sound 
in cascading gauge theory plasma as \cite{aby}
\begin{equation}
v_s^2=\frac 13-\frac 49\ \dd_{cascade}+\calo(\dd_{cascade}^2)\,.
\eqlabel{vsaby}
\end{equation}

In this paper we compute the dispertion relation for the lowest quasinormal mode in the black brane geometry 
holographically dual to thermal cascading gauge theory. At high temperature, $\dd_{cascade}\ll 1$, 
we find (see Eqs.~\eqref{defwwqq}, \eqref{disprel}, \eqref{bv}, \eqref{bgamma})
\begin{equation}
\omega(q)=\frac{1}{\sqrt{3}}\biggl(1-\frac 23\ \dd_{cascade}\biggr)q-i\ \frac{q^2}{6\pi T} 
\biggl(1+\frac 23 \dd_{cascade}\biggr)+\calo\left(\frac{q^3}{T^2}, \dd_{cascade}^2\right)\,,
\eqlabel{mydisp}
\end{equation} 
from which we precisely reproduce the speed of sound extracted from the equation 
of state \eqref{vsaby}. Furthermore, comparing Eqs.~\eqref{dispertion} and \eqref{mydisp} and using the 
universal result \cite{bl1,kss1,bh2}
\begin{equation}
\frac{\eta}{s}=\frac{1}{4\pi}\,,
\eqlabel{etas}
\end{equation} 
we arrive at the 'phenomenological relation' for cascading gauge theory plasma similar to \eqref{proporo}
\begin{equation}
\frac{\zeta}{\eta} = - 2 \, \left(
v_s^2-\frac{1}{3}\right)+\calo\biggl(\dd_{cascade}^2\biggr)\,.
\eqlabel{ratiobulk}
\end{equation}

Most importantly, it appears that phenomenological relation 
\begin{equation}
\frac{\zeta}{\eta} \sim - \calo(1)\times\  \left(
v_s^2-\frac{1}{3}\right)+\calo\left(\biggl[v_s^2-\frac{1}{3}\biggr]^2\right)
\eqlabel{important}
\end{equation}
is a robust prediction for hot  strongly coupled non-conformal gauge theory 
plasma no matter whether scale invariance is broken explicitly (by masses as in Eq.~\eqref{proporo})
or spontaneously (by a strong coupling scale as in Eq.~\eqref{ratiobulk}).    
As such, we expect it to be of relevance to real QCD quark-gluon plasma.
(Note that
the result \eqref{important} appears to disagree with the estimates
$\zeta \sim \eta \left( v_s^2 -1/3\right)^2$
\cite{Hosoya:1983id,Horsley:1985dz}, later criticized in
\cite{Jeon:1995zm}.)

\section{Effective actions and equations  of motion}
Effective 5d action describing supergravity dual to cascading gauge theories is given by \cite{aby}
\begin{equation}
\begin{split}
{S}_5= \frac{1}{16\pi G_5} \int_{\calm_5} \hat{{\rm vol}}_{\calm_5}\
 \biggl\lbrace &
\Omega_1 \Omega_2^4 \biggl(R_{10}-\frac 12 \nabla_\mu \Phi \nabla^\mu
\Phi\biggr)-P^2 \Omega_1 e^{-\Phi}\biggl(
\frac{\nabla_\mu K\nabla^\mu K}{4P^4}+\frac{e^{2\Phi}}{\Omega_1^2}
\biggr)\\
&-\frac 12\frac{K^2}{\Omega_1\Omega_2^4}\biggr\rbrace\,,
\end{split}
\eqlabel{5action}
\end{equation}
where $R_{10}$ is given by 
\begin{equation}
\begin{split}
R_{10}=\hR_5&-2\om_1^{-1}\hg^{\lambda\nu}\biggl(\nabla_{\lambda}\nabla_{\nu}\om_1
\biggr)-8\om_2^{-1}\hg^{\lambda\nu}\biggl(\nabla_{\lambda}\nabla_{\nu}\om_2
\biggr)\\
&-4 \hg^{\lambda\nu}\biggl(2\ \om_1^{-1}\om_2^{-1}\ 
\nabla_\lambda\om_1\nabla_\nu\om_2
+3\ \om_2^{-2}\ \nabla_\lambda\om_2\nabla_{\nu}\om_2\biggr)\\
&+24\ \om_2^{-2}-4\ \om_1^2\ \om_2^{-4}\,,
\end{split}
\eqlabel{ric5}
\end{equation}
with $\hR_5$ being the five dimensional Ricci scalar of the metric 
\begin{equation}
d\hat{s}_{5}^2 =\hg_{\mu\nu}(y) dy^{\mu}dy^{\nu}\,,
\eqlabel{5met}
\end{equation}
and $G_5$ is the five dimensional effective gravitational constant  
\begin{equation}
G_5\equiv \frac{G_{10}}{{\rm vol}_{T^{1,1}}}\,.
\end{equation}

We find it convenient to rewrite the action \eqref{5action} in 5d Einstein frame.
The latter is achieved with the following rescaling
\begin{equation}
\hg_{\mu\nu}\to g_{\mu\nu}\equiv \om_1^{2/3}\om_2^{8/3}\ \hg_{\mu\nu}\,. 
\eqlabel{eing}
\end{equation} 
Further introducing 
\begin{equation}
\om_1=e^{f-4w}\,,\qquad \om_2=e^{f+w}\,,
\eqlabel{defom12}
\end{equation}
the five dimensional effective action becomes 
\begin{equation}
\begin{split}
S_5= \frac{1}{16\pi G_5} \int_{\calm_5} {{\rm vol}}_{\calm_5}\
 \biggl\lbrace &
R_5-\frac{40}{3}(\del f)^2-20(\del w)^2-\frac 12 (\del\Phi)^2-\frac{1}{4P^2}(\del K)^2 e^{-\Phi-4f-4w}\\
&-\calp
\biggr\rbrace\,,
\end{split}
\eqlabel{5actionE}
\end{equation}
where we defined
\begin{equation}
\calp=-24 e^{-\ft {16}{3}f-2w}+4 e^{-\ft {16}{3}f-12w}+P^2 e^{\Phi-\ft {28}{3}f+4w}+\frac 12 K^2 e^{-\ft {40}{3} f }\,.
\eqlabel{calp}
\end{equation}
From Eq.~\eqref{5actionE} we obtain the following equations of motion
\begin{equation}
\begin{split}
0=\square f+\frac{3}{80P^2}e^{-\Phi-4f-4w}(\del K)^2-\frac{3}{80}\ \frac{\del\calp}{\del f}\,,
\end{split}
\eqlabel{eqf}
\end{equation}
\begin{equation}
\begin{split}
0=\square w+\frac{1}{40P^2}e^{-\Phi-4f-4w}(\del K)^2-\frac{1}{40}\ \frac{\del\calp}{\del w}\,,
\end{split}
\eqlabel{eqw}
\end{equation}
\begin{equation}
\begin{split}
0=\square \Phi+\frac{1}{4P^2}e^{-\Phi-4f-4w}(\del K)^2- \frac{\del\calp}{\del \Phi}\,,
\end{split}
\eqlabel{eqphi}
\end{equation}
\begin{equation}
\begin{split}
0=\square K-\del K\del(\Phi+4f+4w)-2P^2 e^{\Phi+4f+4w}\frac{\del\calp}{\del K}
\end{split}
\eqlabel{eqK}
\end{equation}
\begin{equation}
\begin{split}
R_{5\mu\nu}=&\frac{40}{3}\ \del_\mu f\del_\nu f+20\ \del_\mu w\del_\nu w+\frac 12\  \del_\mu \Phi\del_\nu \Phi+\frac{1}{4P^2}
e^{-\Phi-4f-4w}\ \del_\mu K\del_\nu K+\frac 13 g_{\mu\nu}\ \calp\,.
\end{split}
\eqlabel{eqE}
\end{equation}

\subsection{Black brane background geometry}
Taking the  black brane metric ansatz
\begin{equation}
ds_5^2=-c_1^2\ dt^2+c_2^2\ d\vec{x}^2+c_3^2\ dr^2\,,
\eqlabel{bb}
\end{equation}
and assuming that all matter fields $\{f,w,\Phi,K\}$ depend on the radial coordinate $r$ only, 
the background equations of motion are
\begin{equation}
\begin{split}
0=&f''+f'\ \left[\ln\frac{c_1c_2^3}{c_3}\right]'+\frac{3}{80P^2}e^{-\Phi-4f-4w}(K')^2-\frac{3}{80}\ c_3^2\frac{\del\calp}{\del f}\,,
\end{split}
\eqlabel{back1}
\end{equation}
\begin{equation}
\begin{split}
0=&w''+w'\ \left[\ln\frac{c_1c_2^3}{c_3}\right]'+\frac{1}{40P^2}e^{-\Phi-4f-4w}(K')^2-\frac{1}{40}\ c_3^2\frac{\del\calp}{\del w}\,,
\end{split}
\eqlabel{back2}
\end{equation}
\begin{equation}
\begin{split}
0=&\Phi''+\Phi'\ \left[\ln\frac{c_1c_2^3}{c_3}\right]'+\frac{1}{4P^2}e^{-\Phi-4f-4w}(K')^2- c_3^2\frac{\del\calp}{\del \Phi}\,,
\end{split}
\eqlabel{back3}
\end{equation}
\begin{equation}
\begin{split}
0=&K''+K'\ \left[\ln\frac{c_1c_2^3}{c_3}\right]'-K'[\Phi+4f+4w]'-2P^2 c_3^2e^{\Phi+4f+4w}\frac{\del\calp}{\del K}\,,
\end{split}
\eqlabel{back4}
\end{equation}
\begin{equation}
\begin{split}
0=&c_1''+c_1'\ \left[\ln\frac{c_2^3}{c_3}\right]'+\frac 13 c_1c_3^2\calp\,,
\end{split}
\eqlabel{back5}
\end{equation}
\begin{equation}
\begin{split}
0=&c_2''+c_2'\ \left[\ln\frac{c_1c_2^2}{c_3}\right]'+\frac 13 c_2c_3^2\calp\,.
\end{split}
\eqlabel{back6}
\end{equation}
Additionally, there is a first order constraint
\begin{equation}
0=(\Phi')^2+\frac{80}{3}(f')^2+40(w')^2+\frac{1}{2P^2}e^{-\Phi-4f-4w}(K')^2-12[\ln\ c_2]'[\ln\ c_1c_2]'-2c_3^2\calp\,.
\eqlabel{backconst}
\end{equation}

Notice that Eqs.~\eqref{back1}-\eqref{backconst} are equivalent to the set of equations derived in \cite{bhkt2,bhkt3}
provided  we identify
\begin{equation}
c_1\equiv e^{\ft 53f+\hz-3\hx}\,,\qquad c_2\equiv e^{\ft 53f+\hz+\hx}\,,\qquad c_3\equiv e^{\ft 53f-\hz+5\hy}\,,
\qquad f\equiv \hy-\hz\,,
\eqlabel{reltokt}
\end{equation}
where $\{\hx,\hy,\hz\}$ are correspondingly $\{x,y,z\}$ of \cite{bhkt2,bhkt3}.
We would need solution to Eqs.~\eqref{back1}-\eqref{backconst} to leading order in $P^2$. This was originally done in \cite{bhkt3},
but we repeat the main steps to set up conventions. We find convenient to use a new radial coordinate 
\begin{equation}
x\equiv \frac{c_1}{c_2}\,.
\eqlabel{radgauge}
\end{equation}
In terms of $x$ Eqs.~\eqref{back1}-\eqref{backconst} become
\begin{equation}
\begin{split}
0=&c_2''-\frac{5}{c_2}(c_2')^2-\frac 1x c_2'+c_2\biggl\{
\frac{40}{9}(f')^2+\frac{20}{3}(w')^2+\frac 16 (\Phi')^2+\frac{1}{12P^2}e^{-\Phi-4f-4w}(K')^2\biggr\}\,,
\end{split}
\eqlabel{beomx1}
\end{equation}
\begin{equation}
\begin{split}
0=&f''+\frac 1x f'+\frac{3}{80P^2}e^{-\Phi-4f-4w}(K')^2+\frac{\del\ln\calp}{\del f}\ \biggl\{ 
\frac{9}{20}([\ln c_2]')^2+\frac{9}{40x}[\ln c_2]'
-\frac 34 (w')^2\\
&-\frac{3}{160}(\Phi')^2-\frac 12 (f')^2-\frac{3}{320P^2}e^{-\Phi-4f-4w}(K')^2
\biggr\}\,,
\end{split}
\eqlabel{beomx2}
\end{equation}
\begin{equation}
\begin{split}
0=&w''+\frac 1x w'+\frac{1}{40P^2}e^{-\Phi-4f-4w}(K')^2+\frac{\del\ln\calp}{\del w}\ \biggl\{ 
\frac{3}{10}([\ln c_2]')^2+\frac{3}{20x}[\ln c_2]'
-\frac 12 (w')^2\\
&-\frac{1}{80}(\Phi')^2-\frac 13 (f')^2-\frac{1}{160P^2}e^{-\Phi-4f-4w}(K')^2
\biggr\}\,,
\end{split}
\eqlabel{beomx3}
\end{equation}
\begin{equation}
\begin{split}
0=&\Phi''+\frac 1x \Phi'+\frac{1}{4P^2}e^{-\Phi-4f-4w}(K')^2+\frac{\del\ln\calp}{\del \Phi}\ \biggl\{ 
{12}([\ln c_2]')^2+\frac{6}{x}[\ln c_2]'
-20 (w')^2\\
&-\frac{1}{2}(\Phi')^2-\frac {40}{3} (f')^2-\frac{1}{4P^2}e^{-\Phi-4f-4w}(K')^2
\biggr\}\,,
\end{split}
\eqlabel{beomx4}
\end{equation}
\begin{equation}
\begin{split}
0=&K''+[\ln x-\Phi-4f-4w]'K'+\frac{\del\ln\calp}{\del \Phi}\ P^2e^{\Phi+4f+4w}\biggl\{ 
{24}([\ln c_2]')^2+\frac{12}{x}[\ln c_2]'
\\&-40 (w')^2-(\Phi')^2-\frac {80}{3} (f')^2-\frac{1}{2P^2}e^{-\Phi-4f-4w}(K')^2
\biggr\}\,.
\end{split}
\eqlabel{beomx5}
\end{equation}
Notice that Eqs.~\eqref{beomx1}-\eqref{beomx5} have an exact scaling symmetry 
\begin{equation}
c_2\to \lambda\ c_2\,,\qquad f\to f\,,\qquad w\to w\,,\qquad \Phi\to \Phi\,,\qquad K\to K\,,
\eqlabel{symback}
\end{equation}
for a constant $\lambda$.
We will see later that this symmetry has an extension for the fluctuations as well. 
Physically, this symmetry corresponds to choosing a reference energy scale.   

To order $\calo(P^2)$ the solution to Eqs.~\eqref{beomx1}-\eqref{beomx5} takes form \cite{bhkt3}
\begin{equation}
\begin{split}
&c_2=\frac{a}{(1-x^2)^{1/4}}\ \left(1+\frac{P^2}{K_*}\xi(x)\right)\,,\qquad f=-\frac 14\ln\frac{4}{K_*}
+\frac{P^2}{K_*}\eta(x)\,,\\
&w=\frac{P^2}{K_*} \psi(x)\,,\qquad \Phi=\frac{P^2}{K_*}\zeta(x),\qquad  K=K_*+P^2\kappa(x)\,,
\end{split}
\eqlabel{solvebackP2}
\end{equation}
where $a$ is a constant nonextremality parameter, and 
\begin{equation}
\begin{split}
&\xi=\frac{1}{12}(1-\ln(1-x^2))\,,\qquad \kappa=-\frac 12 \ln(1-x^2)\,,\\
&\zeta=\frac{K_*}{P^2}{\Phi_{horizon}} +\frac{\pi^2}{12}-\frac 12\dilog(x)+\frac 12 \dilog(1+x)
-\frac 12 \ln x\ln (1-x)\,,\\
&\eta=-\frac{3(1+x^2)}{80(1-x^2)}\left(\dilog(1-x^2)-\frac {\pi^2}{6}\right)+\frac{1}{20}(1-\ln(1-x^2))\,.
\end{split}
\eqlabel{expP2}
\end{equation}
Furthermore, $\psi$ satisfies the linear differential equation 
\begin{equation}
0=\psi''+\frac 1x \psi'-\frac{3}{(1-x^2)^2}\psi-\frac{1}{10(1-x^2)}\,,
\eqlabel{eqom}
\end{equation}
with the boundary condition
\begin{equation}
\psi=\psi_{horizon}+\calo(x^2)\,,\qquad \psi=-\frac{1}{30}(1-x^2)+\calo\biggl((1-x^2)^{3/2}\biggr)\,,
\eqlabel{asspsi}
\end{equation}
where the second boundary condition will uniquely determine $\psi_{horizon}$.

Finally, the (exact in $P^2$) temperature of the nonextremal solution is given by 
\begin{equation}
(2\pi T)^2=-\frac{\calp c_2^3}{6c_2''}\ \bigg|_{x\to 0_+}\,.
\eqlabel{tempeature}
\end{equation}

\section{Fluctuations}
Now we study  fluctuations in the background geometry
\begin{equation}
\begin{split}
g_{\mu\nu}&\to g_{\mu\nu}+h_{\mu\nu}\,,\\
f&\to f+\dd f\,,\\
w&\to w+\dd w\,,\\
\Phi&\to \Phi+\dd \Phi\,,\\
K&\to K+\dd K\,,
\end{split}
\eqlabel{fluctuations}
\end{equation}
where $\{g_{\mu\nu},f,w,\Phi,K\}$ are the black brane 
background configuration (satisfying Eqs. \eqref{back1}-\eqref{backconst}),
and $\{h_{\mu\nu},\dd f,\dd w,\dd \Phi,\dd K\}$ are the fluctuations. We choose the gauge 
\begin{equation}
h_{tr}=h_{x_ir}=h_{rr}=0\,.
\eqlabel{gaugec}
\end{equation}
 Additionally, 
we assume that all the fluctuations depend only on $(t,x_3,r)$,\ \ie, we have an $O(2)$ rotational symmetry in the 
$x_1x_2$ plane.

At a linearized level we find that the following sets of fluctuations decouple from each other
\begin{equation}
\begin{split}
&\{h_{x_1x_2}\}\,,\\
&\{h_{x_1x_1}-h_{x_2x_2}\}\,,\\
&\{h_{tx_1},\ h_{x_1x_3}\}\,,\\
&\{h_{tx_2},\ h_{x_2x_3}\}\,,\\
&\{h_{tt},\ h_{aa}\equiv h_{x_1x_1}+h_{x_2x_2},\ h_{tx_3},\ h_{x_3x_3},\ \dd f,\ \dd w,\ \dd \Phi,\ \dd K\}\,.
\end{split}
\end{equation}
The last set of fluctuations is a  holographic dual to the sound waves in cascading gauge theory plasma 
which is of interest here. Introduce
\begin{equation}
\begin{split}
h_{tt}=&c_1^2\ \hh_{tt}=e^{-i\w t+iq x_3}\ c_1^2\  H_{tt}\,,\\
h_{tz}=&c_2^2\ \hh_{tz}=e^{-i\w t+iq x_3}\ c_2^2\  H_{tz}\,,\\
h_{aa}=&c_2^2\ \hh_{aa}=e^{-i\w t+iq x_3}\ c_2^2\  H_{aa}\,,\\
h_{zz}=&c_2^2\ \hh_{zz}=e^{-i\w t+iq x_3}\ c_2^2\  H_{zz}\,,\\
\dd f=&e^{-i\w t+iq x_3}\ \calf\,,\\
\dd w=&e^{-i\w t+iq x_3}\ \om\,,\\
\dd \Phi=&e^{-i\w t+iq x_3}\ p\,,\\
\dd K=&e^{-i\w t+iq x_3}\ \k\,,\\
\hh_{ii}=&\hh_{aa}+\hh_{zz}\,,\qquad H_{ii}=H_{aa}+H_{zz}\,,
\end{split}
\eqlabel{rescale}
\end{equation} 
where $\{H_{tt},H_{tz},H_{aa},H_{zz},\calf,\om,p,\k\}$ are functions of a radial coordinate  only. 
Expanding at a linearized level Eqs.~\eqref{eqf}-\eqref{eqE} 
with Eq.~\eqref{fluctuations} and Eq.~\eqref{rescale} we find
the following coupled system of ODE's
\begin{equation}
\begin{split}
0=&H_{tt}''+H_{tt}'\ \left[\ln\frac{c_1^2c_2^3}{c_3}\right]'-H_{ii}'\ [\ln c_1]'
-\frac{c_3^2}{c_1^2}\left(q^2\frac{c_1^2}{c_2^2}\ H_{tt}+\w^2\ H_{ii}+2\w q\ H_{tz}\right)\\
&-\frac 23 c_3^2 \left(\frac{\del\calp}{\del f}\ \calf+\frac{\del\calp}{\del w}\ \om+\frac{\del\calp}{\del \Phi}\ p
+\frac{\del\calp}{\del K}\ \k\right)\,,
\end{split}
\eqlabel{fl1}
\end{equation}
\begin{equation}
\begin{split}
0=&H_{tz}''+H_{tz}'\ \left[\ln\frac{c_2^5}{c_1c_3}\right]'
+\frac{c_3^2}{c_2^2}\ \w q\ H_{aa}\,,
\end{split}
\eqlabel{fl2}
\end{equation}
\begin{equation}
\begin{split}
0=&H_{aa}''+H_{aa}'\ \left[\ln\frac{c_1c_2^5}{c_3}\right]'+(H_{zz}'-H_{tt}')\ [\ln c_2^2]'
+\frac{c_3^2}{c_1^2}\left(\w^2-q^2\frac{c_1^2}{c_2^2}\right)\ H_{aa}\\
&+\frac 43 c_3^2 \left(\frac{\del\calp}{\del f}\ \calf+\frac{\del\calp}{\del w}\ \om+\frac{\del\calp}{\del \Phi}\ p
+\frac{\del\calp}{\del K}\ \k\right)\,,
\end{split}
\eqlabel{fl3}
\end{equation}
\begin{equation}
\begin{split}
0=&H_{zz}''+H_{zz}'\ \left[\ln\frac{c_1c_2^4}{c_3}\right]'+(H_{aa}'-H_{tt}')\ [\ln c_2]'\\
&+\frac{c_3^2}{c_1^2}\left(\w^2\ H_{zz}+2\w q\ H_{tz}+q^2\frac{c_1^2}{c_2^2}(H_{tt}-H_{aa})\right)\\
&+\frac 23 c_3^2 \left(\frac{\del\calp}{\del f}\ \calf+\frac{\del\calp}{\del w}\ \om+\frac{\del\calp}{\del \Phi}\ p
+\frac{\del\calp}{\del K}\ \k\right)\,,
\end{split}
\eqlabel{fl4}
\end{equation}
\begin{equation}
\begin{split}
0=&\calf''+\calf'\ \left[\ln\frac{c_1c_2^3}{c_3}\right]'+\frac 12 f'\ [H_{ii}-H_{tt}]'+\frac{c_3^2}{c_1^2}
\left(\w^2-q^2\frac{c_1^2}{c_2^2}\right)\ \calf\\
&-\frac {3}{80} c_3^2 \biggl(\frac{\del^2\calp}{\del f^2}\ \calf+\frac{\del^2\calp}{\del f\del w}\ \om
+\frac{\del^2\calp}{\del f\del \Phi}\ p+\frac{\del^2\calp}{\del f\del K}\ \k\\
&+\frac{1}{P^2}\ \frac{(K')^2}{c_3^2}e^{-\Phi-4 f-4 w}-\frac{2}{P^2}\ \frac{K'\k'}{c_3^2}e^{-\Phi-4 f-4 w}
(p+4\calf+4\om)
\biggr)\,,
\end{split}
\eqlabel{fl5}
\end{equation}
\begin{equation}
\begin{split}
0=&\om''+\om'\ \left[\ln\frac{c_1c_2^3}{c_3}\right]'+\frac 12 w'\ [H_{ii}-H_{tt}]'+\frac{c_3^2}{c_1^2}
\left(\w^2-q^2\frac{c_1^2}{c_2^2}\right)\ \om\\
&-\frac {1}{40} c_3^2 \biggl(\frac{\del^2\calp}{\del w\del f}\ \calf+\frac{\del^2\calp}{\del w^2}\ \om
+\frac{\del^2\calp}{\del w\del \Phi}\ p+\frac{\del^2\calp}{\del w\del K}\ \k\\
&+\frac{1}{P^2}\ \frac{(K')^2}{c_3^2}e^{-\Phi-4 f-4 w}-\frac{2}{P^2}\ \frac{K'\k'}{c_3^2}e^{-\Phi-4 f-4 w}
(p+4\calf+4\om)\biggr)\,,
\end{split}
\eqlabel{fl6}
\end{equation}
\begin{equation}
\begin{split}
0=&p''+p'\ \left[\ln\frac{c_1c_2^3}{c_3}\right]'+\frac 12 \Phi'\ [H_{ii}-H_{tt}]'+\frac{c_3^2}{c_1^2}
\left(\w^2-q^2\frac{c_1^2}{c_2^2}\right)\ p\\
&-c_3^2 \biggl(\frac{\del^2\calp}{\del \Phi\del f}\ \calf+\frac{\del^2\calp}{\del\Phi\del w}\ \om
+\frac{\del^2\calp}{\del \Phi^2}\ p+\frac{\del^2\calp}{\del \Phi\del K}\ \k\\
&+\frac{1}{4P^2}\ \frac{(K')^2}{c_3^2}e^{-\Phi-4 f-4 w}
(p+4\calf+4\om)-\frac{1}{2P^2}\ \frac{K'\k'}{c_3^2}e^{-\Phi-4 f-4 w}\biggr)\,,
\end{split}
\eqlabel{fl7}
\end{equation}
\begin{equation}
\begin{split}
0=&\k''+\k'\ \left[\ln\frac{c_1c_2^3}{c_3}\right]'+\frac 12 K'\ [H_{ii}-H_{tt}]'+\frac{c_3^2}{c_1^2}
\left(\w^2-q^2\frac{c_1^2}{c_2^2}\right)\ \k\\
&-2P^2e^{\Phi+4f+4w}c_3^2 \biggl(\frac{\del^2\calp}{\del K\del f}\ \calf+\frac{\del^2\calp}{\del K\del w}\ \om
+\frac{\del^2\calp}{\del K\del \Phi}\ p+
\frac{\del^2\calp}{\del K^2}\ \k\\
&+\frac{\del \calp}{\del K}(p+4\calf+4\om)\biggr)-\k'[\Phi+4f+4w]'-K'[p+4\calf+4\om]'\,,
\end{split}
\eqlabel{fl8}
\end{equation}
where all derivatives $\del\calp$ are evaluated on the background geometry. 
Additionally, there are three first order constraints associated with the (partially) fixed diffeomorphism invariance 
\begin{equation}
\begin{split}
0=&\w\left(H_{ii}'+\left[\ln\frac{c_2}{c_1}\right]'\ H_{ii}\right)+q\left(H_{tz}'+2\left[\ln\frac{c_2}{c_1}\right]'\ H_{tz}\right)
\\
&+\w\ \left(\frac{80}{3}f'\calf+40w'\om+\Phi'p+\frac{1}{2P^2}K'\k e^{-\Phi-4f-4w}\right)\,,
\end{split}
\eqlabel{const1}
\end{equation}
\begin{equation}
\begin{split}
0=&q\left(H_{tt}'-\left[\ln\frac{c_2}{c_1}\right]'\ H_{tt}\right)+\frac{c_2^2}{c_1^2}\w\ H_{tz}'-q\ H_{aa}
\\
&-q\ \left(\frac{80}{3}f'\calf+40w'\om+\Phi'p+\frac{1}{2P^2}K'\k e^{-\Phi-4f-4w}\right)\,,
\end{split}
\eqlabel{const2}
\end{equation}
\begin{equation}
\begin{split}
0=&[\ln c_1c_2^2]'H_{ii}'-[\ln{c_2^3}]'\ H_{tt}'+\frac{c_3^2}{c_1^2}
\left(\w^2\ H_{ii}+2\w q\ H_{tz}+q^2\ \frac{c_1^2}{c_2^2}\left(H_{tt}-H_{aa}\right)\right)\\
&+c_3^2\left(\frac{\del\calp}{\del f}\ \calf+\frac{\del\calp}{\del w}\ \om+\frac{\del\calp}{\del \Phi}\ p
+\frac{\del\calp}{\del K}\ \k\right)\\
&-\left(\frac{80}{3}f'\calf'+40w'\om'+\Phi'p'+\frac{1}{2P^2}K'\k' e^{-\Phi-4f-4w}\right)
\\
&+\frac{1}{4P^2}(K')^2e^{-\Phi-4f-4w}(p+4\calf+4\om)\,.
\end{split}
\eqlabel{const3}
\end{equation}
We explicitly verified that Eqs.~\eqref{fl1}-\eqref{fl8} are consistent with constraints 
\eqref{const1}-\eqref{const3}. 

Introducing the gauge invariant fluctuations
\begin{equation}
\begin{split}
Z_H=&4\frac{q}{\w} \ H_{tz}+2\ H_{zz}-H_{aa}\left(1-\frac{q^2}{\w^2}\frac{c_1'c_1}{c_2'c_2}\right)+2\frac{q^2}{\w^2}
\frac{c_1^2}{c_2^2}\ H_{tt}\,,\\
Z_f=&\calf-\frac{f'}{[\ln c_2^4]'}\ H_{aa}\,,\\
Z_w=&\om-\frac{w'}{[\ln c_2^4]'}\ H_{aa}\,,\\
Z_\Phi=&p-\frac{\Phi'}{[\ln c_2^4]'}\ H_{aa}\,,\\
Z_K=&\k-\frac{K'}{[\ln c_2^4]'}\ H_{aa}\,,
\end{split}
\eqlabel{physical}
\end{equation}
we find from Eqs.~\eqref{fl1}-\eqref{fl8}, \eqref{const1}-\eqref{const3}, decoupled
\footnote{To achieve the decoupling 
one has to use the background equations of motion \eqref{back1}-\eqref{backconst}, 
 \ie, the decoupling occurs only on-shell.}
 set of equations 
of motion for $Z$'s
\begin{equation}
\begin{split}
0=&A_HZ_H''+B_HZ_H'+C_HZ_H+D_HZ_f+E_HZ_w+F_HZ_\Phi +G_HZ_K\,, 
\end{split}
\eqlabel{zH}
\end{equation}
\begin{equation}
\begin{split}
0=&A_fZ_f''+B_fZ_f'+C_fZ_H'+D_fZ_H+E_fZ_f+F_fZ_w +G_fZ_\Phi+H_fZ_K'+I_fZ_K\,, 
\end{split}
\eqlabel{zf}
\end{equation}
\begin{equation}
\begin{split}
0=&A_wZ_w''+B_wZ_w'+C_wZ_H'+D_wZ_H+E_wZ_f+F_wZ_w +G_wZ_\Phi+H_wZ_K'+I_wZ_K\,, 
\end{split}
\eqlabel{zw}
\end{equation}
\begin{equation}
\begin{split}
0=&A_\Phi Z_\Phi''+B_\Phi Z_\Phi'+C_\Phi Z_H'+D_\Phi Z_H+E_\Phi Z_f+F_\Phi Z_w +G_\Phi Z_\Phi+H_\Phi Z_K'+I_\Phi Z_K\,, 
\end{split}
\eqlabel{zphi}
\end{equation}
\begin{equation}
\begin{split}
0=&A_K Z_K''+B_K Z_K'+C_K Z_H'+D_K Z_H+E_K Z_f'+F_K Z_f+G_K Z_w'+H_K Z_w\\
&+I_K Z_\Phi'+J_K Z_\Phi+K_K Z_K\,,
\end{split}
\eqlabel{zK}
\end{equation}
where we collected connection coefficients $\{A_{\cdots},\cdots,K_{K} \}$ in 
Appendix \ref{concoeff}. 
Notice that Eqs.~\eqref{zH}-\eqref{zK} have an exact scaling symmetry \eqref{symback} provided the latter is  supplemented with 
\begin{equation}
\w\to\lambda\ \w\,,\qquad q\to \lambda\ q\,,
\eqlabel{supsym}
\end{equation}
while keeping $\{Z_H,Z_f,Z_w,Z_\Phi,Z_K\}$ invariant. 

\section{Hydrodynamic limit, boundary conditions and small $P^2$ expansion}
We study now physical fluctuation equations \eqref{zH}-\eqref{zK} 
in the hydrodynamics approximation, $\w\to 0,\ q\to 0$ with $\ft \w q$ kept constant,
and to leading order in $P^2$. Similar to the $\caln=2^*$ computations \cite{bbs}, 
we would need only leading and next-to-leading (in $q$) solution of \eqref{zH}-\eqref{zK}.  
The computations are greatly simplified with judicial choice of the radial coordinate. Choosing the radial coordinate 
as in \eqref{radgauge}, we find that at the  horizon, $x\to 0_+$, $Z_H\propto x^{\pm i\w/(2\pi T)}$, 
and similarly for $Z_f, Z_w,Z_\Phi,Z_K$. The temperature $T$ is given by Eq.~\eqref{tempeature}.
Incoming boundary conditions on all physical modes implies that 
\begin{equation}
\begin{split}
&Z_H(x)=x^{-i\ww} z_H(x)\,,\qquad Z_f(x)=x^{-i\ww} z_f(x)\,,\qquad Z_w(x)=x^{-i\ww} z_w(x)\,,\\
&Z_\Phi(x)=x^{-i\ww} z_\Phi(x)\,,\qquad Z_K(x)=x^{-i\ww} z_K(x)\,,
\end{split}
\eqlabel{incoming}
\end{equation}
where $\{z_H, z_f, z_w,z_\Phi,z_K\}$ are regular at the horizon; we further introduced 
\begin{equation}
\ww\equiv \frac{\w}{2\pi T}\,,\qquad \qq\equiv\frac{q}{2\pi T} \,.
\eqlabel{defwwqq}
\end{equation}
There is a single integration constant for  these  physical modes, namely, 
the overall scale. Without the loss of generality the latter can be fixed as 
\begin{equation}
z_H(x)\bigg|_{x\to 0_+}=1\,.
\eqlabel{bconditions}
\end{equation}
In this case, the pole dispersion relation is simply determined as \cite{set}
\begin{equation}
z_H(x)\bigg|_{x\to 1_-}=0\,.
\eqlabel{poledisp}
\end{equation}
The other boundary conditions (besides regularity at the horizon and \eqref{poledisp})
are \cite{set}
\begin{equation}
z_f(x)\bigg|_{x\to 1_-}=0\,,\ z_w(x)\bigg|_{x\to 1_-}=0\,,\ z_\Phi(x)\bigg|_{x\to 1_-}=0\,,\ z_K(x)\bigg|_{x\to 1_-}=0\,.
\eqlabel{rembound}
\end{equation}
Let's  introduce 
\begin{equation}
\begin{split}
&z_H=\biggl(z_{H,0}^{0}+P^2\ z_{H,0}^2\biggr)+i\ \qq\  \biggl(z_{H,1}^{0}+P^2\ z_{H,1}^2\biggr)\,,\\
&z_f=\biggl(z_{f,0}^{0}+P^2\ z_{f,0}^2\biggr)+i\ \qq\  \biggl(z_{f,1}^{0}+P^2\ z_{f,1}^2\biggr)\,,\\
&z_w=\biggl(z_{w,0}^{0}+P^2\ z_{w,0}^2\biggr)+i\ \qq\  \biggl(z_{w,1}^{0}+P^2\ z_{w,1}^2\biggr)\,,\\
&z_\Phi=\biggl(z_{\Phi,0}^{0}+P^2\ z_{\Phi,0}^2\biggr)+i\ \qq\  \biggl(z_{\Phi,1}^{0}+P^2\ z_{\Phi,1}^2\biggr)\,,\\
&z_K=\biggl(z_{K,0}^{0}+P^2\ z_{K,0}^2\biggr)+i\ \qq\  \biggl(z_{K,1}^{0}+P^2\ z_{K,1}^2\biggr)\,,
\end{split}
\eqlabel{defzz}
\end{equation}
where the lower index refers to either the leading, $\propto \qq^0$, or to the next-to-leading, $\propto \qq^1$,  
order in the hydrodynamic approximation, and the upper index keeps track of the $P^2$ deformation parameter.   
Additionally, as we are interested in the hydrodynamic pole dispersion relation in the stress-energy correlation functions,
we find it convenient to parameterize 
\begin{equation}
\ww=\frac{\qq}{\sqrt{3}}\biggl(1+\b_v\ P^2\biggr)-i\   \frac {\qq^2}{3}
\biggl(1+\b_\G\ P^2\biggr)\,,
\eqlabel{disprel}
\end{equation}
where the $P^2=0$ coefficients are those of the $\caln=4$ plasma, computed in \cite{ne4}, 
and $\b_v$, $\b_\G$ are constants which are to be determined from the pole dispersion relation
\eqref{poledisp}
\begin{equation}
\begin{split}
&z_{H,0}^2\bigg|_{x\to 1_-}=0,\qquad z_{H,1}^2\bigg|_{x\to 1_-}=0\,.
\end{split}
\eqlabel{poledisp1}
\end{equation}
Using the high-temperature non-extremal cascading gauge theory flow background \eqref{solvebackP2}, parameterizations 
\eqref{defzz}, \eqref{disprel}, we obtain
from Eqs.~\eqref{zH}-\eqref{zK}\footnote{After rewriting them in radial coordinate \eqref{radgauge}. }    
four sets of ODE's describing leading and next-to-leading order in the hydrodynamic approximation and 
$\calo(P^0)$ and $\calo(P^2)$ order in the deformation parameter.

In the remaining part of this section each set is discussed in details.

\subsection{Equations and solution in $\calo(\qq^0)$ and $\calo(P^0)$ order}
To order $\calo(\qq^0)$ and $\calo(P^0)$ we find the following set of equations
\begin{equation}
\begin{split}
0=&[z_{H,0}^0]''+\frac{1-3x^2}{x(1+x^2)}\ [z_{H,0}^0]'+\frac{4}{1+x^2}\ z_{H,0}^0+\frac{32(2(x^3-x)\kappa'+1+x^2)}{(1-x^4)K_*}
\ z_{K,0}^0\,,
\end{split}
\eqlabel{finh00}
\end{equation}
\begin{equation}
\begin{split}
0=&[z_{f,0}^0]''+\frac 1x\ [z_{f,0}^0]'-\frac{8}{(1-x^2)^2}\ z_{f,0}^0+\frac{3\kappa'}{10K_*}\ [z_{K,0}^0]'
+\frac{2}{(1-x^2)^2K_*}\ z_{K,0}^0\,,
\end{split}
\eqlabel{finf00}
\end{equation}
\begin{equation}
\begin{split}
0=&[z_{w,0}^0]''+\frac 1x\ [z_{w,0}^0]'-\frac{3}{(1-x^2)^2}\ z_{w,0}^0+\frac{\kappa'}{5K_*}\ [z_{K,0}^0]'\,,
\end{split}
\eqlabel{finw00}
\end{equation}
\begin{equation}
\begin{split}
0=&[z_{\Phi,0}^0]''+\frac 1x\ [z_{\Phi,0}^0]'+\frac{2\kappa'}{K_*}\ [z_{K,0}^0]'\,,
\end{split}
\eqlabel{finp00}
\end{equation}
\begin{equation}
\begin{split}
0=&[z_{K,0}^0]''+\frac 1x\ [z_{K,0}^0]'\,.
\end{split}
\eqlabel{fink00}
\end{equation}
Notice that at this order, the fluctuations couple only through $z_{K,0}^0$, 
which by itself decouples. 
The most general solution of Eq.~\eqref{fink00} is 
\begin{equation}
z_{K,0}^0=\C_1+\C_2\ \ln x\,.
\eqlabel{sok00}
\end{equation}
Regularity at the horizon and Eq.~\eqref{rembound} imply that 
\begin{equation}
z_{K,0}^0=0\,.
\eqlabel{rk00}
\end{equation}
Vanishing of $z_{K,0}^0=0$ decouples all the remaining fluctuations.
The pattern that we observe here extends to order $\calo(\qq^1)$ in the hydrodynamic approximation 
and to order $\calo(P^2)$ in the deformation parameter: 
\nxt the corresponding gauge invariant fluctuations  couple only through $z_{K,\cdots}^{\cdots}$;
\nxt $z_{K,\cdots}^{\cdots}$ by itself decouples;
\nxt subject to boundary conditions, the unique solution is $z_{K,\cdots}^{\cdots}=0$, 
resulting in decoupling of $\{z_{H,\cdots}^{\cdots},z_{f,\cdots}^{\cdots},z_{w,\cdots}^{\cdots},z_{\Phi,\cdots}^{\cdots}\}$
fluctuations. 
 
Given Eq.~\eqref{rk00}, the remaining equations can be solved analytically. With 
boundary conditions \eqref{bconditions}-\eqref{rembound} unique solutions are 
\begin{equation}
z_{H,0}^0=1-x^2\,,\qquad z_{f,0}^0=z_{w,0}^0=z_{\Phi,0}^0=0\,.
\eqlabel{rest00}
\end{equation}
For $z_{H,0}^0$ we reproduce the sound wave quasinormal mode in the near extremal D3-brane geometry \cite{ne4,bbs}
to leading order in the hydrodynamic approximation.

\subsection{Equations and solution in $\calo(\qq^0)$ and $\calo(P^2)$ order}
Using Eqs.~\eqref{rk00}, \eqref{rest00},  to order $\calo(\qq^0)$ and $\calo(P^2)$ we find the following set of equations
\begin{equation}
\begin{split}
0=&[z_{H,0}^2]''+\frac{1-3x^2}{x(1+x^2)}\ [z_{H,0}^2]'+\frac{4}{1+x^2}\ z_{H,0}^2+\frac{32(2(x^3-x)\kappa'+1+x^2)}{(1-x^4)K_*}
\ z_{K,0}^2\\
&+\frac{8}{3x^2(1+x^2)}\ \biggl(\frac{(\kappa')^2}{K_*}(1-x^2)^2(1+x^2)+6x(1-x^2)^2\xi'+3x^2\b_v\biggr)\,,
\end{split}
\eqlabel{finh02}
\end{equation}
\begin{equation}
\begin{split}
0=&[z_{f,0}^2]''+\frac 1x\ [z_{f,0}^2]'-\frac{8}{(1-x^2)^2}\ z_{f,0}^2+\frac{3\kappa'}{10K_*}\ [z_{K,0}^2]'
+\frac{2}{(1-x^2)^2K_*}\ z_{K,0}^2\\
&+\frac{1}{60(x^2-1)x^3K_*}\ \biggl(40(x^2-1)\eta'+40x(4\eta-\kappa)-7x\biggr)\,,
\end{split}
\eqlabel{finf02}
\end{equation}
\begin{equation}
\begin{split}
0=&[z_{w,0}^2]''+\frac 1x\ [z_{w,0}^2]'-\frac{3}{(1-x^2)^2}\ z_{w,0}^2+\frac{\kappa'}{5K_*}\ [z_{K,0}^2]'\\
&+\frac{1}{30(x^2-1)x^3K_*}\ \biggl(20(x^2-1)\psi'+30x \psi+x\biggr)\,,
\end{split}
\eqlabel{finw02}
\end{equation}
\begin{equation}
\begin{split}
0=&[z_{\Phi,0}^2]''+\frac 1x\ [z_{\Phi,0}^2]'+\frac{2\kappa'}{K_*}\ [z_{K,0}^2]'
+\frac{1}{3(x^2-1)x^3K_*}\ \biggl(2(x^2-1)\zeta'+x\biggr)\,,
\end{split}
\eqlabel{finp02}
\end{equation}
\begin{equation}
\begin{split}
0=&[z_{K,0}^2]''+\frac 1x\ [z_{K,0}^2]'+\frac{2}{3(x^2-1)x^3}\ \biggl((x^2-1)\kappa'+x\biggr)\,.
\end{split}
\eqlabel{fink02}
\end{equation}
Notice that Eqs.~\eqref{finh02}-\eqref{fink02} are equivalent to Eqs.~\eqref{finh00}-\eqref{fink00} of the previous section 
apart from $\calo(P^2)$ sources describing the deformation of the nonextremal cascading gauge theory 
geometry away from the near extremal D3-brane background \eqref{solvebackP2}. The latter is precisely the 
reason that the pattern of coupling of fluctuations is the same as for $P^2=0$.

Using Eq.~\eqref{expP2}, the most general solution of Eq.~\eqref{fink02} is 
\begin{equation}
z_{K,0}^2=\C_1+\C_2\ \ln x\,.
\eqlabel{sok02}
\end{equation}
Regularity at the horizon and Eq.~\eqref{rembound} imply that 
\begin{equation}
z_{K,0}^2=0\,.
\eqlabel{rk02}
\end{equation}
Given Eq.~\eqref{rk02}, the most general solution to Eq.~\eqref{finh02} takes form
\begin{equation}
z_{H,0}^2=\C_1\ \biggl((x^2-1)\ln x-2\biggr)+\C_2\ (1-x^2)-\frac{1}{3K_*}\ \biggl(4+6K_*\b_v\biggr)\,.
\eqlabel{soh02}
\end{equation}
Regularity at the horizon implies that $\C_1=0$, and the boundary condition \eqref{poledisp} determines
\begin{equation}
\b_v=-\frac{2}{3K_*}\,,
\eqlabel{bv}
\end{equation}
in agreement with \cite{aby}. Finally, since $z_{H,0}^0$ already satisfies Eq.~\eqref{bconditions}, we must 
also set $\C_2$ in Eq.~\eqref{soh02} to zero. Thus, we have
\begin{equation}
z_{H,0}^2=0\,.
\eqlabel{rh02}
\end{equation}

Remaining fluctuation equations are discussed in Appendix \ref{fl02}.

\subsection{Equations and solution in $\calo(\qq^1)$ and $\calo(P^0)$ order}
Using Eqs.~\eqref{rk00}, \eqref{rest00},  to order $\calo(\qq^1)$ and $\calo(P^0)$ we find the following set of equations
\begin{equation}
\begin{split}
0=&[z_{H,1}^0]''+\frac{1-3x^2}{x(1+x^2)}\ [z_{H,1}^0]'+\frac{4}{1+x^2}\ z_{H,1}^0+\frac{32(2(x^3-x)\kappa'+1+x^2)}{(1-x^4)K_*}
\ z_{K,1}^0\,,
\end{split}
\eqlabel{finh10}
\end{equation}
\begin{equation}
\begin{split}
0=&[z_{f,1}^0]''+\frac 1x\ [z_{f,1}^0]'-\frac{8}{(1-x^2)^2}\ z_{f,1}^0+\frac{3\kappa'}{10K_*}\ [z_{K,1}^0]'
+\frac{2}{(1-x^2)^2K_*}\ z_{K,1}^0\,,
\end{split}
\eqlabel{finf10}
\end{equation}
\begin{equation}
\begin{split}
0=&[z_{w,1}^0]''+\frac 1x\ [z_{w,1}^0]'-\frac{3}{(1-x^2)^2}\ z_{w,1}^0+\frac{\kappa'}{5K_*}\ [z_{K,1}^0]'\,,
\end{split}
\eqlabel{finw10}
\end{equation}
\begin{equation}
\begin{split}
0=&[z_{\Phi,1}^0]''+\frac 1x\ [z_{\Phi,1}^0]'+\frac{2\kappa'}{K_*}\ [z_{K,1}^0]'\,,
\end{split}
\eqlabel{finp10}
\end{equation}
\begin{equation}
\begin{split}
0=&[z_{K,1}^0]''+\frac 1x\ [z_{K,1}^0]'\,.
\end{split}
\eqlabel{fink10}
\end{equation}
Notice that two sets of equations Eqs.~\eqref{finh00}-\eqref{fink00} and Eqs.~\eqref{finh10}-\eqref{fink10} are 
equivalent.  With 
boundary conditions \eqref{bconditions}-\eqref{rembound} unique solutions are 
\begin{equation}
z_{H,1}^0=z_{f,1}^0=z_{w,1}^0=z_{\Phi,1}^0=z_{K,1}^0=0\,.
\eqlabel{rest10}
\end{equation}
Again, for $z_{H,1}^0$ 
we reproduce the sound wave quasinormal mode in the near extremal D3-brane geometry \cite{ne4,bbs}
to  order $\qq^1$ in the hydrodynamic approximation.

\subsection{Equations and solution in $\calo(\qq^1)$ and $\calo(P^2)$ order}
Using Eqs.~\eqref{rk00}, \eqref{rest00}, \eqref{finf02}-\eqref{finw02}, \eqref{rk02}, \eqref{bv},
\eqref{rh02},  \eqref{rest10}, 
to order $\calo(\qq^1)$ and $\calo(P^2)$ we find the following set of equations
\begin{equation}
\begin{split}
0=&[z_{H,1}^2]''+\frac{1-3x^2}{x(1+x^2)}\ [z_{H,1}^2]'+\frac{4}{1+x^2}\ z_{H,1}^2+\frac{32(2(x^3-x)\kappa'+1+x^2)}{(1-x^4)K_*}
\ z_{K,1}^2\\
&-\frac{8\sqrt{3}}{9K_*\left(x^2+1\right)x^2}\biggl((\kappa')^2\left(1-x^2\right)^2\left((x^2+1)\ln x+3x^2+1\right)\\
&+6x\left(1-x^2\right)^2
\left(\ln x+3\right)\xi'+x^2 \left(3 K_* \beta_\Gamma
-6-2 \ln x\right)\biggr)\,,
\end{split}
\eqlabel{finh12}
\end{equation}
\begin{equation}
\begin{split}
0=&[z_{f,1}^2]''+\frac 1x\ [z_{f,1}^2]'-\frac{8}{(1-x^2)^2}\ z_{f,1}^2+\frac{3\kappa'}{10K_*}\ [z_{K,1}^2]'
+\frac{2}{(1-x^2)^2K_*}\ z_{K,1}^2\\
&-\frac{\sqrt{3}}{60\left(x^2-1\right)^2 \left(x^2+1\right) K_* x^3} \biggl(
40 x^2 K_* \left(1+x^2\right) \left(1-x^2\right)^2 [z_{f,0}^2]'\\
&-\left(1-x^4\right)  
\left(40 \eta' (x^2-1)-40 x \kappa-7 x+160x\eta\right)\biggr)\,,
\end{split}
\eqlabel{finf12}
\end{equation}
\begin{equation}
\begin{split}
0=&[z_{w,1}^2]''+\frac 1x\ [z_{w,1}^2]'-\frac{3}{(1-x^2)^2}\ z_{w,1}^2+\frac{\kappa'}{5K_*}\ [z_{K,1}^2]'\\
&+\frac{\sqrt{3}}{30K_* x^3 (x^4-1)}  \biggl(20 x^2 K_* (1-x^4) [z_{w,0}^2]'\\
&-(x^2+1) \left(20 (x^2-1) \psi'+30 x \psi+x\right)\biggr)\,,
\end{split}
\eqlabel{finw12}
\end{equation}
\begin{equation}
\begin{split}
0=&[z_{\Phi,1}^2]''+\frac 1x\ [z_{\Phi,1}^2]'+\frac{2\kappa'}{K_*}\ [z_{K,1}^2]'
+\frac{\sqrt{3}}{3K_* x^3 (x^4-1)}  \biggl(2 x^2 K_* (1-x^4) [z_{\Phi,0}^2]'\\
&-(x^2+1) \left(2 (x^2-1) \zeta'+x\right)\biggr)\,,
\end{split}
\eqlabel{finp12}
\end{equation}
\begin{equation}
\begin{split}
0=&[z_{K,1}^2]''+\frac 1x\ [z_{K,1}^2]'-\frac{2\sqrt{3}}{9x^3(x^4-1)}  \left((x^2-1) \kappa'+x\right) 
\left((x^2+1)\ln x +3 x^2+1\right)\,.
\end{split}
\eqlabel{fink12}
\end{equation}

Given Eq.~\eqref{expP2}, the most general solution to Eq.~\eqref{fink12} is 
\begin{equation}
z_{K,1}^2=\C_1+\C_2\ \ln x\,.
\eqlabel{sok12}
\end{equation}
Regularity at the horizon and Eq.~\eqref{rembound} imply that 
\begin{equation}
z_{K,1}^2=0\,.
\eqlabel{rk12}
\end{equation}
With Eq.~\eqref{rk12}, the most general solution to Eq.~\eqref{finh12} 
takes the form
\begin{equation}
z_{H,1}^2=\calc_1\ \biggl((x^2-1)\ln x-2\biggr)+\calc_2\ (1-x^2)+\frac{2\sqrt{3}}{9K_*}\biggl(3K_*\b_\Gamma-2\biggr)\,.
\eqlabel{soh12}
\end{equation}
Regularity at the horizon implies that $\C_1=0$, and boundary condition \eqref{poledisp} determines
\begin{equation}
\b_\Gamma=\frac{2}{3K_*}\,.
\eqlabel{bgamma}
\end{equation}
Finally, since $z_{H,0}^0$ already satisfies Eq.~\eqref{bconditions}, we must 
also set $\C_2$ in Eq.~\eqref{soh12} to zero. Thus, we have
\begin{equation}
z_{H,1}^2=0\,.
\eqlabel{rh12}
\end{equation}

Remaining fluctuation equations are discussed in Appendix \ref{fl12}.

\section*{Acknowledgments}
I would like to thank Ofer Aharony  for valuable discussions and  comments on the manuscript. 
I would like to thank Ofer Aharony and Amos Yarom for collaboration on \cite{aby}, which was vital 
for the analysis presented here.
Research at
Perimeter Institute is supported in part by funds from NSERC of
Canada. I gratefully   acknowledge  support by  NSERC Discovery
grant. I would like to thank Aspen Center for Physics for hospitality where part of this 
work was done.

\section*{Appendix}

\appendix

\section{Connection coefficients for Eqs.~\eqref{zH}-\eqref{zK}}\label{concoeff}
\begin{equation}
\begin{split}
A_H=&3 \w^2 c_2' c_2^2 c_3 c_1^2 \biggl(-c_2 c_1 q^2 c_1'-2 c_2' c_1^2 q^2+3 \w^2 c_2' c_2^2\biggr)
\,.\end{split}
\eqlabel{ah}
\end{equation}
\begin{equation}
\begin{split}
B_H=&\w^2 c_2 c_1 \biggl(27 \w^2 c_2^2 c_1 c_2'^3 c_3-42 c_1^3 q^2 c_2'^3 c_3+9 c_2^2 c_1 q^2 c_2' c_1'^2 c_3
-2 c_2^2 c_1^3 q^2 c_2' c_3^3 \calp\\
&+2 c_2^3 c_1^2 q^2 c_1' c_3^3 \calp+3 c_2^2 c_1^2 q^2 c_2' c_1' c_3'
+6 c_2 c_1^3 q^2 c_2'^2 c_3'-3 c_2 c_1^2 q^2 c_2'^2 c_1' c_3-9 \w^2 c_2^3 c_1 c_2'^2 c_3'\\
&+9 \w^2 c_2^3 c_2'^2 c_3 c_1'\biggr)
\,.\end{split}
\eqlabel{bh}
\end{equation}
\begin{equation}
\begin{split}
C_H=&c_3 \w^2 \biggl(-3 \w^2 c_3^2 c_2' c_2^3 q^2 c_1 c_1'+9 \w^4 c_3^2 c_2'^2 c_2^4
-2 c_3^2 q^2 c_1^4 c_2'^2 c_2^2 \calp+36 q^2 c_1^3 c_2'^3 c_1' c_2\\
&-24 q^2 c_1^4 c_2'^4
+4 c_3^2 q^2 c_1^3 c_2' c_1' c_2^3 \calp-2 c_3^2 q^2 c_1^2 c_1'^2 c_2^4 \calp+3 c_3^2 c_2' c_1^3 q^4 c_1' c_2
+6 c_3^2 c_2'^2 c_1^4 q^4\\
&-12 q^2 c_1 c_1'^3 c_2^3 c_2'-15 \w^2 c_3^2 c_2'^2 c_1^2 q^2 c_2^2\biggr)
\,.\end{split}
\eqlabel{ch}
\end{equation}
\begin{equation}
\begin{split}
D_H=&16 c_3 q^2 c_1^2 \biggl(-c_1 c_2'+c_1' c_2\biggr)\biggl(
-\frac{20}{3}  c_2 (-c_1 c_3^2 \calp \omega^2 c_2^2-6 c_2' \omega^2 c_1' c_2+6 c_1 (c_2')^2 \omega^2\\
&+c_1^3 q^2 c_3^2 \calp)f'
-\frac 14 c_1 c_3^2 \frac{\del\calp}{\del f} (2 c_2' c_1^2 q^2-3 c_2' c_2^2 \omega^2+c_1 c_2 c_1' q^2)
\biggr)
\,.\end{split}
\eqlabel{dh}
\end{equation}
\begin{equation}
\begin{split}
E_H=&16 c_3 q^2 c_1^2 \biggl(-c_1 c_2'+c_1' c_2\biggr)\biggl( -10 c_2 (-c_1 c_3^2 \calp \omega^2 c_2^2
-6 c_2' \omega^2 c_1' c_2+6 c_1 c_2'^2 \omega^2\\
&+c_1^3 q^2 c_3^2 \calp) w'
-\frac 14 c_3^2 c_1 \frac{\del\calp}{\del w} (2 c_2' c_1^2 q^2-3 c_2' c_2^2 \omega^2+c_1 c_2 c_1' q^2)
\biggr)
\,.\end{split}
\eqlabel{eh}
\end{equation}
\begin{equation}
\begin{split}
F_H=&16 c_3 q^2 c_1^2 \biggl(-c_1 c_2'+c_1' c_2\biggr)\biggl(
-\frac 14 c_2 (-c_1 c_3^2 \calp \omega^2 c_2^2-6 c_2' \omega^2 c_1' c_2+6 c_1 c_2'^2 \omega^2\\
&+c_1^3 q^2 c_3^2 \calp) 
\Phi'-\frac 14 c_1 c_3^2 \frac{\del\calp}{\del\Phi} (2 c_2' c_1^2 q^2-3 c_2' c_2^2 \omega^2+c_1 c_2 c_1' q^2)
\biggr)
\,.\end{split}
\eqlabel{fh}
\end{equation}
\begin{equation}
\begin{split}
G_H=&16 c_3 q^2 c_1^2 \biggl(-c_1 c_2'+c_1' c_2\biggr)\biggl(
-\frac{1}{8P^2} c_2 e^{-\Phi-4 f-4 w} (-c_1 c_3^2 \calp \omega^2 c_2^2-6 c_2' \omega^2 c_1' c_2
\\
&+6 c_1 c_2'^2 \omega^2+c_1^3 q^2 c_3^2 \calp) K'
-\frac 14 c_3^2 c_1 \frac{\del\calp}{\del K} (2 c_2' c_1^2 q^2-3 c_2' c_2^2 \omega^2+c_1 c_2 c_1' q^2)
\biggr)
\,.\end{split}
\eqlabel{gh}
\end{equation}
\begin{equation}
\begin{split}
A_f=&12 c_2' c_3 c_1^2 c_2^2 \biggl(-2 c_2' c_1^2 q^2+3 c_2' c_2^2 \omega^2
-c_1 c_2 c_1' q^2\biggr)
\,.\end{split}
\eqlabel{af}
\end{equation}
\begin{equation}
\begin{split}
B_f=&12 c_2' c_2 c_1 \biggl(-2 c_2' c_1^2 q^2+3 c_2' c_2^2 \omega^2-c_1 c_2 c_1' q^2\biggr) \biggl(
3 c_3 c_1 c_2'-c_2 c_3' c_1+c_2 c_1' c_3\biggr)
\,.\end{split}
\eqlabel{bf}
\end{equation}
\begin{equation}
\begin{split}
C_f=&-c_3^3 c_1^2 c_2^5 \omega^2 \biggl(\frac{9}{40} c_2' \frac{\del\calp}{\del f}+2 c_2 f' \calp\biggr)
\,.\end{split}
\eqlabel{cf}
\end{equation}
\begin{equation}
\begin{split}
D_f=&-c_3^3 c_1 c_2^4 \omega^2 \biggl(c_1 c_2'-c_2 c_1'\biggr) \biggl(\frac{9}{40} c_2' \frac{\del\calp}{\del f}+2 c_2 f' 
\calp\biggr)
\,.\end{split}
\eqlabel{df}
\end{equation}
\begin{equation}
\begin{split}
E_f=&\frac{9}{20} c_3^3 c_2' c_1^2 c_2^2 \biggl(2 c_2' c_1^2 q^2-3 c_2' c_2^2 \omega^2+c_1 c_2 c_1' q^2\biggr) 
\frac{\del^2\calp}{\del f^2}+4 c_3^3 c_2^3 f' c_1^2 \biggl(-6 c_2' c_2^2 \omega^2\\
&+5 c_2' c_1^2 q^2+c_1 c_2 c_1' q^2\biggr) 
\frac{\del\calp}{\del f}+\frac{320}{3} c_3^3 c_2^4 c_1^2 f'^2 \biggl(-c_2^2 \omega^2+c_1^2 q^2\biggr) \calp\\
&+\frac 35 c_3 c_2' \biggl(\frac{3}{P^2} K'^2 e^{-\Phi-4f-4w} c_1^2 c_2^2-20 c_3^2 c_2^2 \omega^2+20 c_3^2 c_1^2 q^2\biggr) 
\biggl(2 c_2' c_1^2 q^2-3 c_2' c_2^2 \omega^2\\
&+c_1 c_2 c_1' q^2\biggr)
\,.\end{split}
\eqlabel{ef}
\end{equation}
\begin{equation}
\begin{split}
F_f=&\frac{9}{20} c_1^2 c_2^2 c_3^3 c_2' \biggl(2 c_2' c_1^2 q^2-3 c_2' c_2^2 \omega^2+c_1 c_2 c_1' q^2\biggr) 
\frac{\del^2\calp}{\del f\del w}+18 c_2^3 c_3^3 c_1^2 c_2' w' \biggl(-c_2^2 \omega^2\\
&+c_1^2 q^2\biggr) \frac{\del\calp}{\del f}
+4 c_1^2 c_2^3 c_3^3 f' \biggl(2 c_2' c_1^2 q^2-3 c_2' c_2^2 \omega^2+c_1 c_2 c_1' q^2\biggr) \frac{\del\calp}{\del w}\\
&+160 c_1^2 c_2^4 c_3^3 f' w' \biggl(-c_2^2 \omega^2+c_1^2 q^2\biggr) \calp
+\frac {9}{5P^2} c_2^2 c_3 c_1^2 c_2' K'^2 e^{-\Phi-4f-4w} \biggl(2 c_2' c_1^2 q^2\\
&-3 c_2' c_2^2 \omega^2+c_1 c_2 c_1' q^2\biggr)
\,.\end{split}
\eqlabel{ff}
\end{equation}
\begin{equation}
\begin{split}
G_f=&\frac{9}{20} c_1^2 c_2^2 c_3^3 c_2' \biggl(2 c_2' c_1^2 q^2-3 c_2' c_2^2 \omega^2+c_1 c_2 c_1' q^2\biggr) 
\frac{\del^2\calp}{\del f\del\Phi}+\frac{9}{20} c_2^3 c_3^3 c_1^2 c_2' \Phi' \biggl(-c_2^2 \omega^2\\
&+c_1^2 q^2\biggr) 
\frac{\del\calp}{\del f}+4 c_1^2 c_2^3 c_3^3 f' \biggl(2 c_2' c_1^2 q^2-3 c_2' c_2^2 \omega^2+c_1 c_2 c_1' q^2\biggr) 
\frac{\del\calp}{\del\Phi}\\
&+4 c_2^4 c_3^3 c_1^2 f' \Phi' \biggl(-c_2^2 \omega^2+c_1^2 q^2\biggr) \calp
+\frac{9}{20P^2} c_2^2 c_3 c_1^2 c_2' K'^2 e^{-\Phi-4f-4w} \biggl(2 c_2' c_1^2 q^2\\
&-3 c_2' c_2^2 \omega^2+c_1 c_2 c_1' q^2\biggr)
\,.\end{split}
\eqlabel{gf}
\end{equation}
\begin{equation}
\begin{split}
H_f=&-\frac{9}{10P^2} c_1^2 c_2^2 c_3 c_2' K' e^{-\Phi-4f-4w} \biggl(2 c_2' c_1^2 q^2-3 c_2' c_2^2 \omega^2+c_1 c_2 c_1' q^2\biggr)
\,.\end{split}
\eqlabel{hf}
\end{equation}
\begin{equation}
\begin{split}
I_f=&\frac{9}{20} c_3^3 c_1^2 c_2^2 c_2' \biggl(2 c_2' c_1^2 q^2-3 c_2' c_2^2 \omega^2+c_1 c_2 c_1' q^2\biggr) 
\frac{\del^2\calp}{\del f\del K}\\
&+\frac{9}{40P^2} c_2^3 c_3^3 c_1^2 c_2' K' e^{-\Phi-4f-4w} \biggl(
-c_2^2 \omega^2+c_1^2 q^2\biggr) \frac{\del\calp}{\del f}+4 c_2^3 c_3^3 c_1^2 f' \biggl(
2 c_2' c_1^2 q^2\\
&-3 c_2' c_2^2 \omega^2+c_1 c_2 c_1' q^2\biggr) \frac{\del\calp}{\del K}
+\frac{2}{P^2} c_2^4 c_3^3 c_1^2 f' K' e^{-\Phi-4f-4w} \biggl(-c_2^2 \omega^2+c_1^2 q^2\biggr) \calp
\,.\end{split}
\eqlabel{if}
\end{equation}
\begin{equation}
\begin{split}
A_w=&12 c_2' c_3 c_1^2 c_2^2 \biggl(-2 c_2' c_1^2 q^2+3 c_2' c_2^2 \omega^2
-c_1 c_2 c_1' q^2\biggr)
\,.\end{split}
\eqlabel{aw}
\end{equation}
\begin{equation}
\begin{split}
B_w=&12 c_2' c_2 c_1 \biggl(-2 c_2' c_1^2 q^2+3 c_2' c_2^2 \omega^2-c_1 c_2 c_1' q^2\biggr) 
\biggl(3 c_3 c_1 c_2'-c_2 c_3' c_1+c_2 c_1' c_3\biggr)
\,.\end{split}
\eqlabel{bw}
\end{equation}
\begin{equation}
\begin{split}
C_w=&-c_3^3 c_1^2 c_2^5 \omega^2 \biggl(2 c_2 w' \calp+\frac{3}{20} c_2' \frac{\del\calp}{\del w}\biggr)
\,.\end{split}
\eqlabel{cw}
\end{equation}
\begin{equation}
\begin{split}
D_w=&-c_3^3 c_1 c_2^4 \omega^2 \biggl(c_1 c_2'-c_2 c_1'\biggr) \biggl(2 c_2 w' \calp+\frac{3}{20} c_2' \frac{\del\calp}{\del w}
\biggr)
\,.\end{split}
\eqlabel{dw}
\end{equation}
\begin{equation}
\begin{split}
E_w=& \frac{3}{10} c_1^2 c_2^2 c_3^3 c_2' \biggl(2 c_2' c_1^2 q^2-3 c_2' c_2^2 \omega^2+c_1 c_2 c_1' q^2\biggr) 
\frac{\del^2\calp}{\del f\del w}+4 c_3^3 c_1^2 c_2^3 w' \biggl(2 c_2' c_1^2 q^2\\
&-3 c_2' c_2^2 \omega^2+c_1 c_2 c_1' q^2
\biggr) \frac{\del\calp}{\del f}+8 c_3^3 c_1^2 c_2^3 c_2' f' \biggl(-c_2^2 \omega^2+c_1^2 q^2\biggr) 
\frac{\del\calp}{\del w}\\
&+\frac{320}{3} c_1^2 c_2^4 c_3^3 f' w' \biggl(-c_2^2 \omega^2+c_1^2 q^2\biggr) \calp
+\frac {6}{5P^2} c_2^2 c_3 c_1^2 c_2' K'^2 e^{-\Phi-4f-4w} \biggl(2 c_2' c_1^2 q^2\\
&-3 c_2' c_2^2 \omega^2+c_1 c_2 c_1' q^2\biggr)
\,.\end{split}
\eqlabel{ew}
\end{equation}
\begin{equation}
\begin{split}
F_w=&\frac{3}{10} c_3^3 c_2' c_1^2 c_2^2 \biggl(2 c_2' c_1^2 q^2-3 c_2' c_2^2 \omega^2+c_1 c_2 c_1' q^2\biggr) 
\frac{\del^2\calp}{\del w^2}+4 c_3^3 c_2^3 c_1^2 w' \biggl(-6 c_2' c_2^2 \omega^2\\
&+5 c_2' c_1^2 q^2+c_1 c_2 c_1' q^2\biggr) 
\frac{\del\calp}{\del w}+160 c_3^3 c_2^4 c_1^2 w'^2 \biggl(-c_2^2 \omega^2+c_1^2 q^2\biggr) \calp\\
&+
\frac{6}{5} c_3 c_2' \biggl(\frac{1}{P^2}K'^2 e^{-\Phi-4f-4w} c_1^2 c_2^2-10 c_3^2 c_2^2 \omega^2+10 c_3^2 c_1^2 q^2\biggr) 
\biggl(2 c_2' c_1^2 q^2-3 c_2' c_2^2 \omega^2\\
&+c_1 c_2 c_1' q^2\biggr)
\,.\end{split}
\eqlabel{fw}
\end{equation}
\begin{equation}
\begin{split}
G_w=&\frac{3}{10} c_3^3 c_2' c_1^2 c_2^2 \biggl(2 c_2' c_1^2 q^2-3 c_2' c_2^2 \omega^2+c_1 c_2 c_1' q^2\biggr) 
\frac{\del^2\calp}{\del w\del\Phi}+\frac{3}{10} c_1^2 c_2^3 c_3^3 c_2' \Phi' \biggl(
-c_2^2 \omega^2\\
&+c_1^2 q^2\biggr) \frac{\del\calp}{\del w}
+4 c_1^2 c_2^3 c_3^3 w' \biggl(2 c_2' c_1^2 q^2-3 c_2' c_2^2 \omega^2+c_1 c_2 c_1' q^2\biggr) 
\frac{\del\calp}{\del\Phi}\\
&+4 c_1^2 c_2^4 c_3^3 w' \Phi' \biggl(-c_2^2 \omega^2+c_1^2 q^2\biggr) \calp
+\frac{3}{10P^2} c_2^2 c_3 c_1^2 c_2' K'^2 e^{-\Phi-4f-4w} \biggl(2 c_2' c_1^2 q^2\\
&-3 c_2' c_2^2 \omega^2+c_1 c_2 c_1' q^2\biggr)
\,.\end{split}
\eqlabel{gw}
\end{equation}
\begin{equation}
\begin{split}
H_w=&-\frac{3}{5P^2} c_1^2 c_2^2 c_3 c_2' K' e^{-\Phi-4f-4w} \biggl(2 c_2' c_1^2 q^2-3 c_2' c_2^2 \omega^2+c_1 c_2 c_1' q^2\biggr)
\,.\end{split}
\eqlabel{hw}
\end{equation}
\begin{equation}
\begin{split}
I_w=&\frac{3}{10} c_2^2 c_3^3 c_1^2 c_2' \biggl(2 c_2' c_1^2 q^2-3 c_2' c_2^2 \omega^2+c_1 c_2 c_1' q^2\biggr) 
\frac{\del^2\calp}{\del w\del K}\\
&+\frac{3}{20P^2} c_2^3 c_3^3 c_1^2 c_2' K' e^{-\Phi-4f-4w} \biggl(
-c_2^2 \omega^2+c_1^2 q^2\biggr) \frac{\del\calp}{\del w}
+4 c_2^3 c_3^3 c_1^2 w' \biggl(2 c_2' c_1^2 q^2\\
&-3 c_2' c_2^2 \omega^2+c_1 c_2 c_1' q^2\biggr) \frac{\del\calp}{\del K}
+\frac{2}{P^2} c_2^4 c_3^3 c_1^2 w' K' e^{-\Phi-4f-4w} \biggr(-c_2^2 \omega^2+c_1^2 q^2\biggr) \calp
\,.\end{split}
\eqlabel{iw}
\end{equation}
\begin{equation}
\begin{split}
A_\Phi=&12 c_2' c_3 c_1^2 c_2^2 \biggl(-2 c_2' c_1^2 q^2+3 c_2' c_2^2 \omega^2
-c_1 c_2 c_1' q^2\biggr)
\,.\end{split}
\eqlabel{aphi}
\end{equation}
\begin{equation}
\begin{split}
B_\Phi=&12 c_2' c_2 c_1 \biggl(-2 c_2' c_1^2 q^2+3 c_2' c_2^2 \omega^2-c_1 c_2 c_1' q^2\biggr) 
\biggl(3 c_3 c_1 c_2'-c_2 c_3' c_1+c_2 c_1' c_3\biggr)
\,.\end{split}
\eqlabel{bphi}
\end{equation}
\begin{equation}
\begin{split}
C_\Phi=&-c_3^3 c_1^2 c_2^5 \omega^2 \biggl(6 c_2' \frac{\del\calp}{\del\Phi}+2 c_2 \Phi' \calp\biggr)
\,.\end{split}
\eqlabel{cphi}
\end{equation}
\begin{equation}
\begin{split}
D_\Phi=&-c_3^3 c_1 c_2^4 \omega^2 \biggl(c_1 c_2'-c_2 c_1'\biggr) \biggl(6 c_2' \frac{\del\calp}{\del\Phi}+2 c_2 \Phi' 
\calp\biggr)
\,.\end{split}
\eqlabel{dphi}
\end{equation}
\begin{equation}
\begin{split}
E_\Phi=& 12 c_1^2 c_2^2 c_3^3 c_2' \biggl(2 c_2' c_1^2 q^2-3 c_2' c_2^2 \omega^2+c_1 c_2 c_1' q^2\biggr) 
\frac{\del^2\calp}{\del f\del\Phi}+4 c_3^3 c_2^3 c_1^2 \Phi' \biggl(2 c_2' c_1^2 q^2\\
&-3 c_2' c_2^2 \omega^2+c_1 c_2 c_1' q^2\biggr) 
\frac{\del\calp}{\del f}+320 c_3^3 c_2^3 c_1^2 c_2' f' \biggl(-c_2^2 \omega^2+c_1^2 q^2\biggr) \frac{\del\calp}{\del\Phi}\\
&+\frac{320}{3} c_2^4 c_3^3 c_1^2 f' \Phi' \biggl(-c_2^2 \omega^2+c_1^2 q^2\biggr) \calp
+\frac{12}{P^2} c_2^2 c_3 c_1^2 c_2' K'^2 e^{-\Phi-4f-4w} \biggl(2 c_2' c_1^2 q^2\\
&-3 c_2' c_2^2 \omega^2+c_1 c_2 c_1' q^2\biggr)
\,.\end{split}
\eqlabel{ephi}
\end{equation}
\begin{equation}
\begin{split}
F_\Phi=& 12 c_3^3 c_2' c_1^2 c_2^2 \biggl(2 c_2' c_1^2 q^2-3 c_2' c_2^2 \omega^2+c_1 c_2 c_1' q^2\biggr) 
\frac{\del^2\calp}{\del w\del\Phi}+4 c_3^3 c_1^2 c_2^3 \Phi' \biggl(2 c_2' c_1^2 q^2\\
&-3 c_2' c_2^2 \omega^2+c_1 c_2 c_1' q^2\biggr) 
\frac{\del\calp}{\del w}+480 c_3^3 c_1^2 c_2^3 c_2' w' \biggl(-c_2^2 \omega^2+c_1^2 q^2\biggr) \frac{\del\calp}{\del\Phi}\\
&+160 c_1^2 c_2^4 c_3^3 w' \Phi' \biggl(-c_2^2 \omega^2+c_1^2 q^2\biggr) \calp
+\frac{12}{P^2} c_2^2 c_3 c_1^2 c_2' K'^2 e^{-\Phi-4f-4w} \biggl(2 c_2' c_1^2 q^2\\
&-3 c_2' c_2^2 \omega^2+c_1 c_2 c_1' q^2\biggr)
\,.\end{split}
\eqlabel{fphi}
\end{equation}
\begin{equation}
\begin{split}
G_\Phi=&12 c_3^3 c_2' c_1^2 c_2^2 \biggl(2 c_2' c_1^2 q^2-3 c_2' c_2^2 \omega^2+c_1 c_2 c_1' q^2\biggr) 
\frac{\del^2\calp}{\del\Phi^2}+4 c_3^3 c_2^3 \Phi' c_1^2 \biggl(-6 c_2' c_2^2 \omega^2\\
&+5 c_2' c_1^2 q^2+c_1 c_2 c_1' q^2\biggr) 
\frac{\del\calp}{\del\Phi}+4 c_3^3 c_2^4 c_1^2 \Phi'^2 \biggl(-c_2^2 \omega^2+c_1^2 q^2\biggr) \calp\\
&+3 c_3 c_2' \biggl(\frac{1}{P^2}K'^2 e^{-\Phi-4f-4w} c_1^2 c_2^2-4 c_3^2 c_2^2 \omega^2+4 c_3^2 c_1^2 q^2\biggr) 
\biggl(2 c_2' c_1^2 q^2-3 c_2' c_2^2 \omega^2\\
&+c_1 c_2 c_1' q^2\biggr)
\,.\end{split}
\eqlabel{gphi}
\end{equation}
\begin{equation}
\begin{split}
H_\Phi=&-\frac{6}{P^2} c_1^2 c_2^2 c_3 c_2' K' e^{-\Phi-4f-4w} \biggl(2 c_2' c_1^2 q^2-3 c_2' c_2^2 \omega^2+c_1 c_2 c_1' 
q^2\biggr)
\,.\end{split}
\eqlabel{hphi}
\end{equation}
\begin{equation}
\begin{split}
I_\Phi=&12 c_2^2 c_3^3 c_1^2 c_2' \biggl(2 c_2' c_1^2 q^2-3 c_2' c_2^2 \omega^2+c_1 c_2 c_1' q^2\biggr) 
\frac{\del^2\calp}{\del \Phi\del K}\\
&+\frac{6}{P^2} c_2^3 c_3^3 c_1^2 c_2' K' e^{-\Phi-4f-4w} \biggl(-c_2^2 \omega^2+c_1^2 q^2\biggr) 
\frac{\del\calp}{\del\Phi}+4 c_2^3 c_3^3 c_1^2 \Phi' \biggl(2 c_2' c_1^2 q^2\\
&-3 c_2' c_2^2 \omega^2+c_1 c_2 c_1' q^2\biggr) 
\frac{\del\calp}{\del K}+\frac{2}{P^2} c_2^4 c_3^3 c_1^2 \Phi' K' e^{-\Phi-4f-4w} \biggl(-c_2^2 \omega^2+c_1^2 q^2\biggr) \calp
\,.\end{split}
\eqlabel{iphi}
\end{equation}
\begin{equation}
\begin{split}
A_K=&12 c_2' c_3 c_1^2 c_2^2 \biggl(-2 c_2' c_1^2 q^2+3 c_2' c_2^2 \omega^2
-c_1 c_2 c_1' q^2\biggr)
\,.\end{split}
\eqlabel{ak}
\end{equation}
\begin{equation}
\begin{split}
B_K=&12 c_2' c_2 c_1 \biggl(-2 c_2' c_1^2 q^2+3 c_2' c_2^2 \omega^2-c_1 c_2 c_1' q^2\biggr) 
\biggl(3 c_3 c_1 c_2'-c_2 c_3' c_1+c_2 c_1' c_3\biggr)\\
&+12 c_3 c_2' c_1^2 c_2^2 \biggl(4 w'+\Phi'+4 f'\biggr) 
\biggl(2 c_2' c_1^2 q^2-3 c_2' c_2^2 \omega^2+c_1 c_2 c_1' q^2\biggr)
\,.\end{split}
\eqlabel{bk}
\end{equation}
\begin{equation}
\begin{split}
C_K=&-c_3^3 c_1^2 c_2^5 \omega^2 \biggl(12 c_2' P^2 \frac{\del\calp}{\del K} e^{\Phi+4f+4w}+2 c_2 K' \calp\biggr)
\,.\end{split}
\eqlabel{ck}
\end{equation}
\begin{equation}
\begin{split}
D_K=&-c_3^3 c_1 c_2^4 \omega^2 \biggl(c_1 c_2'-c_2 c_1'\biggr) \biggl(
12 c_2' P^2 \frac{\del\calp}{\del K} e^{\Phi+4f+4w}+2 c_2 K' \calp\biggr)
\,.\end{split}
\eqlabel{dk}
\end{equation}
\begin{equation}
\begin{split}
E_K=&48 c_3 c_2' K' c_1^2 c_2^2 \biggl(2 c_2' c_1^2 q^2-3 c_2' c_2^2 \omega^2+c_1 c_2 c_1' q^2\biggr)
\,.\end{split}
\eqlabel{ek}
\end{equation}
\begin{equation}
\begin{split}
F_K=&24 c_3^3 c_1^2 c_2^2 c_2' P^2 e^{\Phi+4f+4w} \biggl(2 c_2' c_1^2 q^2-3 c_2' c_2^2 \omega^2+c_1 c_2 c_1' q^2\biggr) 
\frac{\del^2\calp}{\del f\del K}\\
&+4 c_3^3 c_1^2 c_2^3 K' \biggl(2 c_2' c_1^2 q^2-3 c_2' c_2^2 \omega^2+c_1 c_2 c_1' q^2\biggr) 
\frac{\del\calp}{\del f}\\
&+32 c_3^3 c_1^2 c_2^2 c_2' P^2 e^{\Phi+4f+4w} \biggl(20 c_1^2 c_2 q^2 f'-20 c_2^3 \omega^2 f'-9 c_2' 
c_2^2 \omega^2+3 c_1 c_2 c_1' q^2\\
&+6 c_2' c_1^2 q^2\biggr) \frac{\del\calp}{\del K}+\frac{320}{3} 
c_3^3 c_1^2 c_2^4 K' f' \biggl(-c_2^2 \omega^2+c_1^2 q^2\biggr) \calp
\,.\end{split}
\eqlabel{fk}
\end{equation}
\begin{equation}
\begin{split}
G_K=&48 c_3 c_2' K' c_1^2 c_2^2 \biggl(2 c_2' c_1^2 q^2-3 c_2' c_2^2 \omega^2+c_1 c_2 c_1' q^2\biggr)
\,.\end{split}
\eqlabel{gk}
\end{equation}
\begin{equation}
\begin{split}
H_K=&24 c_3^3 c_2^2 c_1^2 c_2' P^2 e^{\Phi+4f+4w} \biggl(2 c_2' c_1^2 q^2-3 c_2' c_2^2 \omega^2+c_1 c_2 c_1' q^2\biggr) 
\frac{\del^2\calp}{\del w\del K}\\
&+4 c_3^3 c_2^3 c_1^2 K' \biggl(2 c_2' c_1^2 q^2-3 c_2' c_2^2 \omega^2+c_1 c_2 c_1' q^2\biggr)
 \frac{\del\calp}{\del w}\\
&+96 c_3^3 c_2^2 c_1^2 c_2' P^2 e^{\Phi+4f+4w} \biggl(10 c_1^2 c_2 q^2 w'+2 c_2' c_1^2 q^2
-10 c_2^3 \omega^2 w'-3 c_2' c_2^2 \omega^2\\
&+c_1 c_2 c_1' q^2\biggr) \frac{\del\calp}{\del K}
+160 c_3^3 c_2^4 c_1^2 K' w' \biggl(-c_2^2 \omega^2+c_1^2 q^2\biggr) \calp
\,.\end{split}
\eqlabel{hk}
\end{equation}
\begin{equation}
\begin{split}
I_K=&12 c_3 c_2' K' c_1^2 c_2^2 \biggl(2 c_2' c_1^2 q^2-3 c_2' c_2^2 \omega^2+c_1 c_2 c_1' q^2\biggr)
\,.\end{split}
\eqlabel{ik}
\end{equation}
\begin{equation}
\begin{split}
J_K=&24 c_3^3 c_2^2 c_1^2 c_2' P^2 e^{\Phi+4f+4w} \biggl(2 c_2' c_1^2 q^2-3 c_2' c_2^2 \omega^2+c_1 c_2 c_1' q^2\biggr) 
\frac{\del^2\calp}{\del \Phi\del K}\\
&+4 c_3^3 c_2^3 c_1^2 K' \biggl(2 c_2' c_1^2 q^2-3 c_2' c_2^2 \omega^2+c_1 c_2 c_1' q^2\biggr) 
\frac{\del\calp}{\del\Phi}\\
&+24 c_3^3 c_2^2 c_1^2 c_2' P^2 e^{\Phi+4f+4w} \biggl(c_1 c_2 c_1' q^2-c_2^3 \omega^2 \Phi'+c_1^2 c_2 q^2 
\Phi'+2 c_2' c_1^2 q^2\\
&-3 c_2' c_2^2 \omega^2\biggr) \frac{\del\calp}{\del K}+4 c_3^3 c_2^4 c_1^2 K' \Phi' \biggl(
-c_2^2 \omega^2+c_1^2 q^2\biggr) \calp
\,.\end{split}
\eqlabel{jk}
\end{equation}
\begin{equation}
\begin{split}
K_K=&24 c_3^3 c_2^2 c_1^2 c_2' P^2 e^{\Phi+4f+4w} \biggl(2 c_2' c_1^2 q^2-3 c_2' c_2^2 \omega^2+c_1 c_2 c_1' q^2\biggr) 
\frac{\del^2\calp}{\del K^2}\\
&+4 c_3^3 c_2^3 K' c_1^2 \biggl(-6 c_2' c_2^2 \omega^2+5 c_2' c_1^2 q^2+c_1 c_2 c_1' q^2\biggr) 
\frac{\del\calp}{\del K}\\
&+\frac{2}{P^2} 
c_3^3 c_2^4 c_1^2 K'^2 e^{-\Phi-4f-4w} \biggl(-c_2^2 \omega^2+c_1^2 q^2\biggr) \calp
\\
&+12 c_2' c_3^3 \biggl(-c_2^2 \omega^2+c_1^2 q^2\biggr) \biggl(2 c_2' c_1^2 q^2-3 c_2' c_2^2 \omega^2+c_1 c_2 c_1' q^2\biggr)
\,.\end{split}
\eqlabel{kk}
\end{equation}

\section{Matter fluctuations at order $\calo(\qq^0)$  and $\calo(P^2)$}\label{fl02}
Unique solution to Eq.~\eqref{finp02} subject to regularity at the horizon and the boundary condition
\eqref{rembound} is 
\begin{equation}
z_{\Phi,0}^2=\frac{1-x^2}{12 K_* x^2}\ \ln(1-x^2)\,.
\eqlabel{rp02}
\end{equation} 

Eqs.~\eqref{finf02}, \eqref{finw02} are second order linear non-homogeneous ordinary differential equations which general 
homogeneous solution can be found analytically. Thus it is straightforward  to write down a formal solution to
\eqref{finf02}, \eqref{finw02} in quadratures. We do not need these explicit expressions for the 
computation of the hydrodynamic properties of the high-temperature cascading gauge theory plasma, so we only argue 
here that regularity at the horizon and the Dirichlet  condition at the boundary determine two integration constants. 
We verified numerically that these are the two {\it independent} integration constants. The latter implies that 
$z_{f,0}^2, z_{w,0}^2$ are uniquely determined.   

Consider first Eq.~\eqref{finf02}. Its general solution near the horizon $x\to 0_+$ and the 
boundary $(1-x^2)\equiv y\to 0_+$ takes form
\begin{equation}
\begin{split}
z_{f,0}^2=&\calc_1\ (2+\ln x)+\calc_2+\calo(x^2\ln x)\,,
\end{split}
\eqlabel{f02h}
\end{equation}  
\begin{equation}
\begin{split}
z_{f,0}^2=&\frac 1y\ \left(-2\hc_2-4\hc_1-\frac{3}{40K_*}\right)+\left(\hc_2+2\hc_1+\frac{3}{80K_*}\right)
-\frac{1}{160K_*}\ y+\calo(y^2\ln y)\,,
\end{split}
\eqlabel{f02bb}
\end{equation}  
where $\hc_1=\hc_1(\calc_1,\calc_2)$ and $\hc_2=\hc_2(\calc_1,\calc_2)$.
Regularity at the horizon implies that $\calc_1=0$; Dirichlet condition at the boundary further constraints 
$\calc_2$:
\begin{equation}
0=\hc_2+2\hc_1+\frac{3}{80K_*}=\hc_2\left(0,\calc_2\right)+2\hc_1\left(0,\calc_2\right)+\frac{3}{80K_*}\,.
\eqlabel{f02con}
\end{equation}
We verified numerically that Eq.~\eqref{f02con} indeed has a solution.

Similarly, the general solution near the horizon $x\to 0_+$ and the 
boundary $(1-x^2)\equiv y\to 0_+$ of Eq.~\eqref{finw02} takes form
\begin{equation}
\begin{split}
z_{w,0}^2=&\calc_1\ \ln x+\calc_2+\calo(x^2\ln x)\,,
\end{split}
\eqlabel{w02h}
\end{equation}  
\begin{equation}
\begin{split}
K_*\ z_{w,0}^2=&\frac {1}{{y}^{1/2}}\ \hc_1-\frac 14 y^{1/2}\ \hc_1-\frac{1}{90} y+y^{3/2}
\left(\hc_2-\frac{1}{32}\ \hc_1\ \ln y\right)+\calo\left(y^2\right) \,,
\end{split}
\eqlabel{w02bb}
\end{equation}  
where we used the power series solution for $\psi$ (see Eqs.~\eqref{eqom}, \eqref{asspsi}) near the horizon
\begin{equation}
\psi=\psi_{horizon}+x^2\left(\frac 34\ \psi_{horizon}+\frac{1}{40}\right)
+\calo(x^4)\,,
\eqlabel{psih}
\end{equation}
and at the boundary 
\begin{equation}
\psi=-\frac {1}{30}\ y+ \hpsi\ y^{3/2}-\frac{2}{75}\ y^2+\calo\left(y^{5/2}\right)\,.
\eqlabel{psib}
\end{equation}
Parameters $\psi_{horizon}$, 
$\hpsi=\hpsi(\phi_{horizon})$ are uniquely determined from the second boundary condition in Eq.~\eqref{asspsi}.
In Eq.~\eqref{w02bb}  $\hc_1=\hc_1(\calc_1,\calc_2)$ and $\hc_2=\hc_2(\calc_1,\calc_2)$.
Regularity at the horizon implies that $\calc_1=0$; Dirichlet condition at the boundary further constraints 
$\calc_2$:
\begin{equation}
0=\hc_1=\hc_1\left(0,\calc_2\right)\,.
\eqlabel{w02con}
\end{equation}
We verified numerically that Eq.~\eqref{w02con} indeed has a solution.

\section{Matter fluctuations at order $\calo(\qq^1)$  and $\calo(P^2)$}\label{fl12}
Unique solution to Eq.~\eqref{finp12} subject to regularity at the horizon and the boundary condition
\eqref{rembound} is 
\begin{equation}
z_{\Phi,1}^2=-\frac{\sqrt{3}}{36K_*x^2}\biggl(2 x^2\left(\ln x-2\right)\ln(1-x^2)+x^2\dilog(x^2)+4\ln(1-x^2)\biggr)\,.
\eqlabel{rp12}
\end{equation} 

As in Appendix \ref{fl02}, here we only argue that  $z_{f,1}^2, z_{w,1}^2$ solutions to Eqs.~\eqref{finf12} and
\eqref{finw12} with appropriate 
boundary conditions exist, and are unique.  

Consider first Eq.~\eqref{finf12}. Its general solution near the horizon $x\to 0_+$ and the 
boundary $(1-x^2)\equiv y\to 0_+$ takes form
\begin{equation}
\begin{split}
z_{f,1}^2=&\calc_1\ (2+\ln x)+\calc_2+\calo(x^2\ln x)\,,
\end{split}
\eqlabel{f12h}
\end{equation}  
\begin{equation}
\begin{split}
z_{f,1}^2=&\frac 1y\ \left(-2\hc_2-4\hc_1\right)+\left(\hc_2+2\hc_1\right)
+\frac{\sqrt{3}}{160K_*}\ y+\calo(y^2\ln y)\,,
\end{split}
\eqlabel{f12b}
\end{equation}  
where we used power series solution at the horizon (boundary) for $z_{f,0}^2$.
Regularity at the horizon implies that $\calc_1=0$; Dirichlet condition at the boundary further constraints 
$\calc_2$:
\begin{equation}
0=\hc_2+2\hc_1=\hc_2\left(0,\calc_2\right)+2\hc_1\left(0,\calc_2\right)\,.
\eqlabel{f12con}
\end{equation}
We verified numerically that Eq.~\eqref{f12con} indeed has a solution.

Finally, the general solution near the horizon $x\to 0_+$ and the 
boundary $(1-x^2)\equiv y\to 0_+$ of Eq.~\eqref{finw12} takes form
\begin{equation}
\begin{split}
z_{w,1}^2=&\calc_1\ \ln x+\calc_2+\calo(x^2\ln x)\,,
\end{split}
\eqlabel{w12h}
\end{equation}  
\begin{equation}
\begin{split}
K_*\ \times\  z_{w,1}^2=&\frac {1}{{y}^{1/2}}\ \hc_1-\frac 14 y^{1/2}\ \hc_1+\frac{\sqrt{3}}{90} y+y^{3/2}
\left(\hc_2-\frac{1}{32}\ \hc_1\ \ln y\right)+\calo\left(y^2\right)\,, 
\end{split}
\eqlabel{w12bb}
\end{equation}  
where we used the power series solutions for $\psi$ and $z_{w,0}^2$ near the horizon (boundary).
Regularity at the horizon implies that $\calc_1=0$; Dirichlet condition at the boundary further constraints 
$\calc_2$:
\begin{equation}
0=\hc_1=\hc_1\left(0,\calc_2\right)\,.
\eqlabel{w12con}
\end{equation}
We verified numerically that Eq.~\eqref{w12con} indeed has a solution.


\begin{thebibliography}{99}

\bibitem{m9711}J.~M.~Maldacena,
``The large $N$ limit of superconformal field theories and supergravity,''
Adv.\ Theor.\ Math.\ Phys.\  {\bf 2}, 231 (1998)
[Int.\ J.\ Theor.\ Phys.\  {\bf 38}, 1113 (1999)]
[arXiv:hep-th/9711200].


\bibitem{m2}
  O.~Aharony, S.~S.~Gubser, J.~M.~Maldacena, H.~Ooguri and Y.~Oz,
  Phys.\ Rept.\  {\bf 323}, 183 (2000)
  [arXiv:hep-th/9905111].



\bibitem{ss}
D.~T.~Son and A.~O.~Starinets,
``Minkowski-space correlators in AdS/CFT correspondence: Recipe and
applications,''
JHEP {\bf 0209}, 042 (2002)
[arXiv:hep-th/0205051].


\bibitem{hs}
C.~P.~Herzog and D.~T.~Son,
``Schwinger-Keldysh propagators from AdS/CFT correspondence,''
JHEP {\bf 0303}, 046 (2003)
[arXiv:hep-th/0212072].


\bibitem{ne1}
G.~Policastro, D.~T.~Son and A.~O.~Starinets,
``The shear viscosity of strongly coupled $N = 4$
supersymmetric Yang-Mills  plasma,''
Phys.\ Rev.\ Lett.\  {\bf 87}, 081601 (2001) [arXiv:hep-th/0104066].

\bibitem{ne2}
G.~Policastro, D.~T.~Son and A.~O.~Starinets,
``From AdS/CFT correspondence to hydrodynamics,''
JHEP {\bf 0209}, 043 (2002) [arXiv:hep-th/0205052].

\bibitem{ne3}
C.~P.~Herzog,
``The hydrodynamics of M-theory,''
JHEP {\bf 0212}, 026 (2002) [arXiv:hep-th/0210126].

\bibitem{ne4}
G.~Policastro, D.~T.~Son and A.~O.~Starinets,
``From AdS/CFT correspondence to hydrodynamics. II: Sound waves,''
JHEP {\bf 0212}, 054 (2002) [arXiv:hep-th/0210220].

\bibitem{ne5}
C.~P.~Herzog,
``The sound of M-theory,''
Phys.\ Rev.\ D {\bf 68}, 024013 (2003) [arXiv:hep-th/0302086].


\bibitem{kss}
P.~Kovtun, D.~T.~Son and A.~O.~Starinets,
``Holography and hydrodynamics: Diffusion on stretched horizons,''
JHEP {\bf 0310}, 064 (2003)
[arXiv:hep-th/0309213].


\bibitem{bl1}
  A.~Buchel and J.~T.~Liu,
  ``Universality of the shear viscosity in supergravity,''
  Phys.\ Rev.\ Lett.\  {\bf 93}, 090602 (2004)
  [arXiv:hep-th/0311175].




\bibitem{kss1}
  P.~Kovtun, D.~T.~Son and A.~O.~Starinets,
  ``Viscosity in strongly interacting quantum field theories from black hole
  physics,''
  Phys.\ Rev.\ Lett.\  {\bf 94}, 111601 (2005)
  [arXiv:hep-th/0405231].


\bibitem{bh1}
  A.~Buchel,
  ``N = 2* hydrodynamics,''
  Nucl.\ Phys.\ B {\bf 708}, 451 (2005)
  [arXiv:hep-th/0406200].




  

\bibitem{bls}
  A.~Buchel, J.~T.~Liu and A.~O.~Starinets,
  ``Coupling constant dependence of the shear viscosity in N = 4 supersymmetric
  Yang-Mills theory,''
  Nucl.\ Phys.\ B {\bf 707}, 56 (2005)
  [arXiv:hep-th/0406264].

\bibitem{bh2}
  A.~Buchel,
  ``On universality of stress-energy tensor correlation functions in
  supergravity,''
  Phys.\ Lett.\ B {\bf 609}, 392 (2005)
  [arXiv:hep-th/0408095].


\bibitem{set}
  P.~K.~Kovtun and A.~O.~Starinets,
  ``Quasinormal modes and holography,''
  arXiv:hep-th/0506184.



\bibitem{bbs}
  P.~Benincasa, A.~Buchel and A.~O.~Starinets,
  ``Sound waves in strongly coupled non-conformal gauge theory plasma,''
  arXiv:hep-th/0507026.
 



\bibitem{r1}
D.~Teaney,
``Effect of shear viscosity on spectra, elliptic flow, and Hanbury Brown-Twiss
radii,''
Phys.\ Rev.\ C {\bf 68}, 034913 (2003).


\bibitem{r2}
E.~Shuryak,
``Why does the quark gluon plasma at RHIC behave as a nearly ideal fluid?,''
Prog.\ Part.\ Nucl.\ Phys.\  {\bf 53}, 273 (2004)
[arXiv:hep-ph/0312227].

\bibitem{r3}
D.~Molnar and M.~Gyulassy,
``Saturation of elliptic flow at RHIC: Results from the covariant elastic
parton cascade model MPC,''
Nucl.\ Phys.\ A {\bf 697}, 495 (2002)
[Erratum-ibid.\ A {\bf 703}, 893 (2002)]
[arXiv:nucl-th/0104073].

\bibitem{w}
  R.~Donagi and E.~Witten,
  ``Supersymmetric Yang-Mills Theory And Integrable Systems,''
  Nucl.\ Phys.\ B {\bf 460}, 299 (1996)
  [arXiv:hep-th/9510101].



\bibitem{pw}
K.~Pilch and N.~P.~Warner,
``N = 2 supersymmetric RG flows and the IIB dilaton,''
Nucl.\ Phys.\ B {\bf 594}, 209 (2001)
[arXiv:hep-th/0004063].


\bibitem{bpp}
  A.~Buchel, A.~W.~Peet and J.~Polchinski,
  ``Gauge dual and noncommutative extension of an N = 2 supergravity
  solution,''
  Phys.\ Rev.\ D {\bf 63}, 044009 (2001)
  [arXiv:hep-th/0008076].

\bibitem{j}
  N.~J.~Evans, C.~V.~Johnson and M.~Petrini,
  ``The enhancon and N = 2 gauge theory/gravity RG flows,''
  JHEP {\bf 0010}, 022 (2000)
  [arXiv:hep-th/0008081].



\bibitem{bl}
A.~Buchel and J.~T.~Liu,
``Thermodynamics of the N = 2* flow,''
JHEP {\bf 0311}, 031 (2003)
[arXiv:hep-th/0305064].


\bibitem{kn}
  I.~R.~Klebanov and N.~A.~Nekrasov,
  ``Gravity duals of fractional branes and logarithmic RG flow,''
  Nucl.\ Phys.\ B {\bf 574}, 263 (2000)
  [arXiv:hep-th/9911096].


\bibitem{kt}
  I.~R.~Klebanov and A.~A.~Tseytlin,
  ``Gravity duals of supersymmetric SU(N) x SU(N+M) gauge theories,''
  Nucl.\ Phys.\ B {\bf 578}, 123 (2000)
  [arXiv:hep-th/0002159].


\bibitem{ks}
  I.~R.~Klebanov and M.~J.~Strassler,
  ``Supergravity and a confining gauge theory: Duality cascades and
  chiSB-resolution of naked singularities,''
  JHEP {\bf 0008}, 052 (2000)
  [arXiv:hep-th/0007191].

\bibitem{ktreview}
  M.~J.~Strassler,
  ``The duality cascade,''
  arXiv:hep-th/0505153.



\bibitem{bhkt1}
  A.~Buchel,
  ``Finite temperature resolution of the Klebanov-Tseytlin singularity,''
  Nucl.\ Phys.\ B {\bf 600}, 219 (2001)
  [arXiv:hep-th/0011146].

\bibitem{bhkt2}
A.~Buchel, C.~P.~Herzog, I.~R.~Klebanov, L.~A.~Pando Zayas and A.~A.~Tseytlin,
``Non-extremal gravity duals for fractional D3-branes on the conifold,''
JHEP {\bf 0104}, 033 (2001)
[arXiv:hep-th/0102105].

\bibitem{bhkt3}
S.~S.~Gubser, C.~P.~Herzog, I.~R.~Klebanov and A.~A.~Tseytlin,
``Restoration of chiral symmetry: A supergravity perspective,''
JHEP {\bf 0105}, 028 (2001)
[arXiv:hep-th/0102172].

\bibitem{aby}
  O.~Aharony, A.~Buchel and A.~Yarom,
  ``Holographic renormalization of cascading gauge theories,''
  arXiv:hep-th/0506002.


\bibitem{qu1}
  A.~Parnachev and A.~Starinets,
  ``The silence of the little strings,''
  arXiv:hep-th/0506144.

\bibitem{qu2}
  A.~Buchel,
  ``A holographic perspective on Gubser-Mitra conjecture,''
  arXiv:hep-th/0507275.

\bibitem{Hosoya:1983id}
  A.~Hosoya, M.~a.~Sakagami and M.~Takao,
  ``Nonequilibrium Thermodynamics In Field Theory: Transport Coefficients,''
  Annals Phys.\  {\bf 154}, 229 (1984).

\bibitem{Horsley:1985dz}
  R.~Horsley and W.~Schoenmaker,
  ``Quantum Field Theories Out Of Thermal Equilibrium. 1. General
  Considerations,''
  Nucl.\ Phys.\ B {\bf 280}, 716 (1987).



\bibitem{Jeon:1995zm}
  S.~Jeon and L.~G.~Yaffe,
  ``From Quantum Field Theory to Hydrodynamics: Transport Coefficients and
  Effective Kinetic Theory,''
  Phys.\ Rev.\ D {\bf 53}, 5799 (1996)
  [arXiv:hep-ph/9512263].


\bibitem{kw}
  I.~R.~Klebanov and E.~Witten,
  ``Superconformal field theory on threebranes at a Calabi-Yau  singularity,''
  Nucl.\ Phys.\ B {\bf 536}, 199 (1998)
  [arXiv:hep-th/9807080].





\end{thebibliography}
\end{document}